\documentclass[twocolumn,amsthm]{autart}

\usepackage[T1]{fontenc}
\usepackage{amsmath,amssymb}
\usepackage{graphicx}
\usepackage[caption=false]{subfig}
\usepackage{xcolor}
\usepackage{tikz}
\usepackage{pgfplots}
\usetikzlibrary{arrows.meta}
\pgfplotsset{compat=1.18}

\newtheorem{theorem}{Theorem}
\newtheorem{lemma}{Lemma}
\newtheorem{proposition}{Proposition}
\theoremstyle{definition}
\newtheorem{assumption}{Assumption}

\newtheorem{remark}{Remark}
\newtheorem{example}{Example}

\begin{document}
\begin{frontmatter}
\title{Toward a decentralized design framework for discrete-time distributed algorithms via their emergent dynamics\thanksref{funding}}

\thanks[funding]{This work supported by the National Research Foundation of Korea (NRF) grant funded by the Korea government (MSIT) (No. RS-2026-25504174). Corresponding author: Jin Gyu Lee.}

\author[snu]{Jinwook Heo}\ead{jwheo@cdsl.kr},
\author[snu]{Hyungbo Shim}\ead{hshim@snu.ac.kr},
\author[snu]{Jin Gyu Lee}\ead{jingyu.lee@snu.ac.kr}
\address[snu]{ASRI, Department of Electrical and Computer Engineering, Seoul National University, Seoul, Republic of Korea}


\begin{abstract}
This paper develops a procedure for designing discrete-time distributed algorithms to achieve a desired collective behavior. Each agent combines its node and coupling updates at every iteration, in either explicit or implicit form. For this structure, we derive necessary and sufficient conditions on the local parameters for realizing a desired continuous-time blended dynamics. These conditions yield a decentralized procedure for computing the parameters. A Tikhonov-type approximation result relates the resulting agent trajectories to the blended dynamics on a rescaled time axis. Under a stability condition imposed only on the blended dynamics, we further establish uniform-in-time approximation bounds. We also characterize the trade-off between approximation accuracy and the iteration count guaranteed by these bounds. The framework is illustrated through four applications in distributed computation and estimation.
\end{abstract}
\end{frontmatter}

\section{Introduction}

    Distributed algorithms enable networked agents to coordinate and compute using only locally available information, without relying on a central unit. For heterogeneous agents, it is useful to understand the collective behavior that emerges from their individual dynamics and interactions. A simpler description of this collective behavior can also provide a basis for designing the local updates of the agents.

    For continuous-time heterogeneous agents, the approach developed in \cite{JGLEE_tool,JKim} describes the collective behavior that emerges when the agents approximately synchronize under sufficiently strong coupling. The continuous-time system describing this behavior is called the \emph{blended dynamics} and has been used as a tool for designing distributed algorithms. For digital implementation, the discrete-time counterpart of the blended dynamics approach was proposed in~\cite{JWKim}. Its multi-step structure is closely related to discrete-time implementations of high-gain designs~\cite{LWang}, where repeated coupling updates emulate the effect of strong coupling. In particular, \cite{JWKim} proposes a method in which each agent~$i\in\mathcal{N}:=\{1,\ldots,N\}$ runs its node update $f_i$ every $K$-th iteration and runs the coupling update otherwise:
    \begin{equation*}
    x_i[n+1]
    =
    \begin{cases}
    f_i(x_i[n]),
    & \text{if } \operatorname{mod}(n,K)=0,\\
    \sum_{j=1}^N \mathsf w_{ij}x_j[n],
    & \text{otherwise},
    \end{cases}
    \end{equation*}
    where $x_i[n]$ is the state and $\mathsf w_{ij}$ is the weight used in the coupling update. For sufficiently large $K$, the method yields approximate synchronization. However, because all agents must run the node update synchronously at the same iterations, the method is vulnerable to asynchronous perturbations, as illustrated in Section~\ref{subsection:page}.

    We instead propose a discrete-time structure in which the node and coupling updates are conducted simultaneously at every iteration. Specifically, the explicit and implicit updates of agent~$i\in\mathcal{N}$ are given by
    \begin{subequations}\label{eq:simplest}
    \begin{equation}\label{eq:simplest_explict}
    \begin{split}
     x_i[n+1] = \frac{a_i}{k}f_i(x_i[n])+b_i x_i[n] + c_i\sum_{j\in\mathcal N_i}\alpha_{ij}x_j[n],
    \end{split}
    \end{equation}
    \begin{equation}\label{eq:simplest_implict}
    \begin{split}
    x_i[n+1] = \frac{a_i}{k}f_i(x_i[n+1])+b_i x_i[n] + c_i\sum_{j\in\mathcal N_i}\alpha_{ij}x_j[n],
    \end{split}
    \end{equation}
    \end{subequations}
    where $x_i[n]\in \mathbb{R}^m$ is the state, $f_i: \mathbb{R}^m \to \mathbb{R}^m$ is a globally Lipschitz node-update function, $\mathcal{N}_i$ is the set of agents sending information to agent~$i$, and $\alpha_{ij}$ is the coupling weight from agent~$j$ to agent~$i$. The scalars $a_i$, $b_i$, and $c_i$ are design parameters. The common parameter $k>0$ corresponds to $K$ in the multi-step method introduced above but need not be an integer. In particular, the factor $1/k$ uniformly spreads the effect of the node update over all iterations, rather than applying it impulsively only every $K$-th iteration. The two forms \eqref{eq:simplest_explict} and \eqref{eq:simplest_implict} differ only in the iteration index used in the argument of $f_i$: $x_i[n]$ is used in the explicit form and $x_i[n+1]$ is used in the implicit form.

    Restricting our attention to the linearly parameterized structure \eqref{eq:simplest}, we first analyze its limiting behavior as $k\to \infty$ in Section~\ref{section:2}. We show that the limiting behavior exhibits synchronization with individually scaled states and that the synchronized trajectory is governed by a continuous-time system describing the emergent behavior of the network. Following the terminology of the continuous-time approach~\cite{JGLEE_tool,JKim}, we continue to refer to this system as the blended dynamics. Under a stability condition on the blended dynamics, we further show that, for every sufficiently large but finite $k$, each agent's state eventually remains close to the corresponding trajectory of the blended dynamics over an infinite time horizon. This is analogous to the discrete-time singular perturbation analysis given in~\cite{Khalil}. 

    Based on this analysis, we then turn to the design problem and take the following dynamics as the desired collective behavior:
    \[
    \dot{\mathsf{s}}(t) =  \sum_{i=1}^N w_i f_i(v_i\mathsf{s}(t)),
    \]
    where $\mathsf{s}(t)\in\mathbb{R}^m$, and $w_i>0$ and $v_i>0$ are the desired weights and scaling factors, respectively. Our objective is to choose the local parameters $a_i$, $b_i$, and $c_i$ so that \eqref{eq:simplest} generates this prescribed blended dynamics. In Section~\ref{section:design}, we derive necessary and sufficient conditions under which the desired blended dynamics can be realized by \eqref{eq:simplest}, and then develop a decentralized procedure for computing the corresponding parameters using local information.

    We emphasize that, although the desired blended dynamics is formulated in continuous time, it serves only as an auxiliary design target for the discrete-time algorithm \eqref{eq:simplest}. Since the desired blended dynamics is independent of $k$, it provides a common reference for comparing the algorithm for different values of $k$. 
    Increasing $k$ reduces the approximation bounds, but the factor $1/k$ also reduces the effect of the node update at each iteration and may increase the number of iterations needed to achieve a prescribed accuracy. Section~\ref{subsection:trade-off} analyzes this trade-off and gives conditions under which the iteration count guaranteed by these bounds is minimized at a finite value of $k$.

    In the discrete-time setting, existing studies on synchronization of heterogeneous multi-agent systems have mostly considered linear agent models and relied on structural assumptions such as a common internal model or simultaneous stabilizability to achieve asymptotic synchronization~\cite{DZhang,SLi,XXu,YYan}. Our interest is fundamentally different from these prior works in that we consider nonlinear node dynamics and study approximate synchronization without imposing such structural assumptions. The proposed framework also maintains useful features of the blended dynamics approaches~\cite{JGLEE_tool,JKim,JWKim}. The stability condition is imposed only on the blended dynamics, so the individual node dynamics need not be stable. Moreover, the algorithms do not require specific initial conditions. This initialization-free property is useful for plug-and-play operation, where agents may join or leave the network.

    The general strategy of prescribing desired collective behavior through the blended dynamics has been used in continuous-time distributed computation and optimization~\cite{JGLEE_design,SLee}, decentralized observers~\cite{TKim}, and initialization-free economic dispatch~\cite{HYun}. In this paper, the proposed distributed algorithm design is illustrated through four applications in Section~\ref{section:applications}: network size estimation, estimation of the reciprocal of each entry of the positive left eigenvector of the Laplacian matrix, a weighted least-squares solver, and PageRank estimation. Simulations of selected applications are presented in Section~\ref{sec:simulations}. In particular, in the asynchronous setting considered in Section~\ref{subsection:page}, the proposed PageRank estimates remain close to the desired scores, whereas those of the multi-step method in~\cite{JWKim} do not. Proofs of the main results are provided in the Appendix.

    \textbf{Note for reading:} Throughout the paper, whenever equations or statements are given in paired forms labeled (a) and (b), they denote the explicit and implicit cases, respectively, while the label without (a) or (b) refers to both cases. For brevity, the notation $[\,\cdot\,]$ indicates evaluation at $n$ in the explicit case and at $n+1$ in the implicit case.
    
\subsection{Notation and preliminaries on graph theory} \label{subsec:notation}

Let $1_N$ and $0_N$ be the $N\times1$ all-ones and all-zeros vectors, respectively, and let $I_N$ denote the $N\times N$ identity matrix. For vectors $a_i$, $i\in\mathcal{N}$, define $\operatorname{col}(a_i)_{i\in\mathcal{N}} := [a_1^\top,\ldots,a_N^\top]^\top$. For scalars $x_i$, $i \in \mathcal{N}$, define $\operatorname{diag}(x_i)_{i \in \mathcal{N}}:=\operatorname{diag}(x_1,\ldots,x_N)$. For a vector $x$ and a matrix $A$, $\|x\|$ and $\|A\|$ denote the Euclidean norm and the induced matrix $2$-norm, respectively. Let $\mathbb{R}_{>0}^N$ be the set of $N$-dimensional vectors with positive entries. A vector or matrix is called positive (nonnegative) if all its entries are positive (nonnegative). Inequalities between vectors are understood componentwise. For a real square matrix $\mathsf{M}$, $\rho(\mathsf{M})$ denotes the spectral radius of $\mathsf{M}$. The Kronecker product is denoted by $\otimes$, and we write $\mathsf{M}_{\otimes m}:=\mathsf{M}\otimes I_m$ for simplicity. For $a\in\mathbb{R}$, define $(a)_+:=\max\{a,0\}$.

A digraph $\mathcal{G}=(\mathcal{N},\mathcal{E}, \mathcal{A})$ with $N\ge2$ has a node set $\mathcal{N}=\{1,\ldots,N\}$ and an edge set $\mathcal{E}\subseteq\mathcal{N}\times\mathcal{N}$, where $(i,j)\in\mathcal{E}$ denotes an edge from $i$ to $j$. We assume no self-loops, i.e., $(i,i)\notin\mathcal{E}$. The in-neighbor set of node $i$ is $\mathcal{N}_i:=\{j\in\mathcal{N}:(j,i)\in\mathcal{E}\}$, and $\mathcal{G}$ is strongly connected if there exists a directed path between any ordered pair of nodes. The weighted adjacency matrix $\mathcal{A}=[\alpha_{ij}]\in\mathbb{R}^{N\times N}$ is defined by $\alpha_{ij}>0$ if $(j,i)\in\mathcal{E}$ and $\alpha_{ij}=0$ otherwise. The weighted in-degree and out-degree of node $i$ are $d_i:=\sum_{j=1}^N \alpha_{ij}$ and $d_i^{\text{out}}:=\sum_{j=1}^N \alpha_{ji}$, respectively. The Laplacian matrix is $\mathcal{L}:=\mathcal{D}-\mathcal{A}$, where $\mathcal{D}=\operatorname{diag}(d_i)_{i \in \mathcal{N}}$. Then $\mathcal{L} 1_N=0_N$, and all eigenvalues of $\mathcal{L}$ have nonnegative real parts. If $\mathcal{G}$ is strongly connected, the zero eigenvalue is simple and the normalized left eigenvector $\zeta=\operatorname{col}(\zeta_i)_{i \in \mathcal{N}}\in\mathbb{R}^N_{>0}$ satisfies $\zeta^\top\mathcal{L}=0_N^\top$ and $\zeta^\top 1_N=1$.  In particular, if $\mathcal{G}$ is undirected and connected, then $\zeta=\frac1N 1_N$. For $N\ge2$, $\mathcal{A}$ is irreducible if and only if $\mathcal{G}$ is strongly connected. A nonnegative matrix $\mathsf{M}\in\mathbb{R}^{N\times N}$ is \emph{primitive} if there exists $m\in\mathbb{N}$ such that $\mathsf{M}^m$ is positive; in particular, any irreducible nonnegative matrix with at least one positive diagonal entry is primitive. We use the following Perron--Frobenius theorem for primitive matrices~\cite{Bullo}.

   \begin{proposition} \label{lem_Perron}
        Let $\mathsf{M}\in\mathbb{R}^{N\times N}$ be a primitive matrix, and let its eigenvalues be ordered as $|\lambda_1|\ge|\lambda_2|\ge\cdots\ge|\lambda_N|$. Then $\rho(\mathsf{M})=\lambda_1>0$ is a simple eigenvalue and satisfies $\lambda_1 > |\lambda_j|$ for all $j=2,\ldots,N$. Moreover, there exist positive right and left eigenvectors $p$ and $q$ of $\lambda_1$, respectively, such that
        $\mathsf{M}p=\lambda_1 p$, $q^\top \mathsf{M}=\lambda_1 q^\top$, and $q^\top p=1$.
        In particular, no other eigenvalue admits a positive right or left eigenvector.

     \end{proposition}

\section{Analysis of Emergent Collective Behaviors} \label{section:2}

In this section, we analyze how \eqref{eq:simplest} behaves as $k\to \infty$ and identify the condition under which the agents exhibit collective behavior. A key result is that, when viewed on an appropriately rescaled time axis, this behavior is described by the continuous-time blended dynamics. Under an additional stability condition on the blended dynamics, we further obtain a uniform convergence result.

\subsection{Limiting behavior of \eqref{eq:simplest} as $k\to\infty$}

    We assume that the communication graph $\mathcal{G}$ induced by the weights $\alpha_{ij}$ satisfies the following.
 
   \begin{assumption} \label{assum:graph}
     The graph $\mathcal{G}$ is strongly connected.     
    \end{assumption}

    Let us define $x[n] := \operatorname{col}(x_i[n])_{i\in\mathcal{N}}$, $F(x):=\operatorname{col}(f_i(x_i))_{i \in \mathcal{N}}$, $A:=\operatorname{diag}(a_i)_{i\in\mathcal{N}}$, $B:=\operatorname{diag}(b_i)_{i\in\mathcal{N}}$, and $C:=\operatorname{diag}(c_i)_{i\in\mathcal{N}}$. Then, \eqref{eq:simplest} can be compactly written as
	\begin{align}\label{eq:general_stack_implicit}
	x[n+1] &= \frac{A_{\otimes m}}{k} F({x[\, \cdot \,]}) +  \mathcal{W}_{\otimes m} x[n],
	\end{align} 
    where $x_i[\,\cdot\,]$ represents $x_i[n]$ in the explicit case and $x_i[n+1]$ in the implicit case. Just for readability, one may simply consider the scalar case $m=1$, where the Kronecker products can be ignored. We impose the following assumption on the parameters $b_i$ and $c_i$.

    \begin{assumption} \label{assum:b,c} 
    The parameters satisfy $b_i\ge 0$ and $c_i>0$ for all $i\in\mathcal{N}$, and $b_i>0$ for at least one $i\in\mathcal{N}$.
    \end{assumption}

    This assumption can be understood from a graph-theoretic perspective. Since $c_i>0$ for all $i\in\mathcal{N}$, the matrix $C\mathcal{A}$ can be regarded as a weighted adjacency matrix of the same graph. The condition that $b_i>0$ for at least one $i\in\mathcal{N}$ adds a self-loop to $\mathcal{W}:=B+C\mathcal{A}$. 
    
    Together with Assumption~\ref{assum:graph}, these properties guarantee that $\mathcal{W}$ is primitive.

  \begin{assumption} \label{assum:W}
      The spectral radius of $\mathcal{W}$ equals one, i.e., $\rho(\mathcal{W})=1$.
  \end{assumption}

    \begin{remark}
    Let us briefly comment on the other two cases, $\rho(\mathcal{W})<1$ and $\rho(\mathcal{W})>1$. If $\rho(\mathcal{W})<1$, the limiting agent states approach zero as $k\to\infty$, yielding only trivial behavior. If $\rho(\mathcal{W})>1$, the network can become unstable for sufficiently large $k$, and the states may diverge. We therefore focus on $\rho(\mathcal{W})=1$ to study the collective behavior described below. Further details are provided in Appendix~\ref{app:spectral-cases}.
    \end{remark}
    
    Under Assumptions~\ref{assum:graph}--\ref{assum:W}, the matrix $\mathcal{W}$ is primitive and satisfies $\rho(\mathcal{W})=1$. Thus, applying Proposition~\ref{lem_Perron} to $\mathcal{W}$, there exist $R, Z \in \mathbb{R}^{N  \times (N-1)}$ such that
      \[
      \mathcal{W} = \begin{bmatrix} p & {R} \end{bmatrix} \begin{bmatrix} 1 & 0_{N-1}^\top \\ 0_{N-1} & {\Lambda}\end{bmatrix} \begin{bmatrix} q^\top \\ {Z}^{\top}\end{bmatrix},
      \]
      where ${Z}^\top {R} = I_{N-1}$, and ${Z}^{\top}p = {R}^{\top} q = 0_{N-1}$, and ${\Lambda} \in \mathbb{R}^{(N-1) \times (N-1)}$ has eigenvalues $\lambda_2,\ldots,\lambda_N$. Now, the coordinate change
    \begin{align} \label{eq:coordinate_change}
        \begin{bmatrix}
         \xi_1 \\ \tilde{\xi}    
        \end{bmatrix} = 
        \begin{bmatrix}
             q^\top_{\otimes m} \\  {Z}^{\top}_{\otimes m}
        \end{bmatrix} {x},
    \end{align}    
    decomposes the state into a weighted sum $\xi_1\in\mathbb{R}^m$ and a disagreement state $\tilde{\xi}\in\mathbb{R}^{m(N-1)}$, so that $x = p_{\otimes m}\xi_1 + R_{\otimes m} \tilde{\xi}$. Then, the system \eqref{eq:general_stack_implicit} can be represented as
	\begin{equation}
    \begin{aligned}\label{eq:xi_1_dynamics}
        & \xi_1[n+1] \\ & =  \xi_1[n] + \frac{1}{k}  \sum_{i=1}^N q_i a_i f_i(p_i\xi_1[\, \cdot \,] + (R_i\otimes I_m)\tilde{\xi}[\, \cdot \,]) ,
    \end{aligned}
    \end{equation}
    \begin{equation}
    \begin{aligned} \label{eq:xi_tilde_dynamics}
    	& \tilde{\xi}[n+1] \\ & =  {\Lambda}_{\otimes m} \tilde{\xi}[n]  + \frac{({Z}^\top A)_{\otimes m}}{k} F( p_{\otimes m}\xi_1[\, \cdot \,] + R_{\otimes m}\tilde{\xi}[\, \cdot \,]),
	\end{aligned}
    \end{equation}
   where $p_i$ and $q_i$ are the $i$-th elements of $p$ and $q$, respectively, and ${R}_i$ is the $i$-th row of ${R}$. Since $\rho({\Lambda})=|\lambda_2|<1$, the linear part of the disagreement dynamics is stable. The remaining nonlinear term is weakened by $1/k$.

    This suggests that, for sufficiently large $k$, $\tilde{\xi}$ becomes arbitrarily small, and hence $\xi_1$ approximately follows the reduced dynamics obtained by letting $\tilde{\xi}=0_{m(N-1)}$ in~\eqref{eq:xi_1_dynamics}. Denoting the state of this reduced dynamics by $s[n]\in\mathbb{R}^m$, we obtain
    \begin{subequations} \label{eq:emer_dyn}
       \begin{align}  \label{eq:emer_dyn_explicit}
            s[n+1] &=   s[n] + \frac{1}{k}  \sum_{i=1}^N {q_i a_i} f_i(p_is[n]),\\
             \label{eq:emer_dyn_implicit}
                s[n+1] &=  s[n] + \frac{1}{k}  \sum_{i=1}^N {q_i a_i} f_i(p_is[n+1]). 
        \end{align}
    \end{subequations}
  The resulting structure \eqref{eq:emer_dyn} closely resembles the explicit and implicit Euler discretizations, respectively, of a continuous-time system with sampling time
    \[
        h_k := \frac{\tau_{\mathsf{s}}}{k}.
    \]  
    As this sampling time converges to zero as $k \to \infty$, the standard Euler interpretation dictates that the limiting behavior can simply be described by the continuous-time system 
    \begin{align}\label{eq:cont_emer}
	   \dot{\mathsf{s}}(t) = \sum_{i=1}^N \frac{q_i a_i}{\tau_{\mathsf{s}}} f_i(p_i \mathsf{s} (t))=: \frac{1}{\tau_{\mathsf{s}}} f_{\mathsf{s}} (\mathsf{s}(t)),
	\end{align}      
    where $\mathsf{s}(t) \in \mathbb{R}^m$. We refer to \eqref{eq:cont_emer} as the {\it blended dynamics} of \eqref{eq:simplest}, which is independent of $k$. This continuous-time blended dynamics is the quasi-steady-state subsystem in the discrete-time singular perturbation theory described in~\cite{Khalil}. Here, $\tau_{\mathsf{s}} > 0$ is a time-scale parameter introduced for generality, and its role in the design will be explained in Section~\ref{section:design}.

   For a given continuous-time instant $T>0$, define the corresponding discrete-time index by
    \begin{align} \label{eq:time_step}
      n_{T,k}:=\left\lceil{\frac{T}{h_k}} \right\rfloor,
    \end{align}
    where $\lceil\cdot\rfloor$ denotes the rounding operator. 
    Let $x_i[n,k]$ and $s[n,k]$ denote the trajectories of \eqref{eq:simplest} and \eqref{eq:emer_dyn}, respectively ($x_i[0,k]=x_i[0]$, $s[0,k]=s[0]$). Since $n_{T,k}h_k\to T$ as $k\to\infty$, this allows us to compare $x_i[n_{T,k},k]$ and $s[n_{T,k},k]$ with $\mathsf{s}(T)$ for each fixed continuous-time instant $T$.
    These observations are justified by the following results.

    \begin{lemma} \label{lem:discret_error}
    For any $\mathsf{s}(0)\in \mathbb{R}^m$, the solution $s[\cdot,k]$ of \eqref{eq:emer_dyn} and the solution $\mathsf{s}(\cdot)$ of \eqref{eq:cont_emer} with $s[0]=\mathsf{s}(0)$ satisfy, for each $T>0$,
        \begin{align} \label{eq:discretize_error}
             \lim_{k\to \infty}  s[n_{T,k},k] = \mathsf{s}(T). 
        \end{align}
    \end{lemma}
    
    Building upon this standard Euler approximation result, we further extend it to our distributed algorithm as follows. This can be interpreted as a discrete-time Tikhonov-type theorem for the multi-agent system \eqref{eq:simplest}. In this interpretation, the continuous-time blended dynamics \eqref{eq:cont_emer} is the corresponding quasi-steady-state subsystem, with $\epsilon:=1/k$ being the small singular perturbation parameter. 

  \begin{theorem} \label{thm:lim_sol}
    Let Assumptions \ref{assum:graph}--\ref{assum:W} hold. For any $x_i[0]\in \mathbb{R}^{m}$, $i \in \mathcal{N}$, the solution $x_i[\cdot,k]$ of \eqref{eq:simplest} and the solution $\mathsf{s}(\cdot)$ of \eqref{eq:cont_emer} with $\mathsf{s}(0)=\sum_{i=1}^N q_i x_i[0]$ satisfy, for each $T>0$,
     \begin{align} \label{eq:lim_sol_CT}
    \lim_{k\to \infty}  {x}_i\left[n_{T,k},k\right] = p_i\mathsf{s} (T),\qquad \forall i \in \mathcal{N}.   
     \end{align}%
  \end{theorem}%
Theorem~\ref{thm:lim_sol} shows that the agents approach an individually scaled synchronized state (with scaling $p_i$) and that the synchronized trajectory follows the continuous-time blended dynamics \eqref{eq:cont_emer}. However, the convergence results in \eqref{eq:discretize_error} and \eqref{eq:lim_sol_CT} are only pointwise in the continuous-time instant $T$. By imposing the additional stability condition introduced in the next subsection, both convergences become {uniform in time}.\footnote{We say that the convergence is uniform in time if, for each $\epsilon>0$, there exist $T_0>0$ and $k_0>0$ such that, for each $k>k_0$, $\|x_i[n_{T,k},k]-p_i\mathsf{s}(T)\|\le\epsilon$, $\forall i\in\mathcal{N}$, $\forall T\ge T_0$. This conclusion is established at the end of Appendix~\ref{app:proof-theorem2}.} This conclusion is consistent with the discrete-time singular perturbation theory~\cite{Khalil} and has a direct analogue in the continuous-time blended dynamics theory~\cite{JKim,JGLEE_tool}.

\subsection{Uniform convergence under stability of \eqref{eq:cont_emer}} \label{subsection:uniform}

    To guarantee the convergence over an infinite time horizon for a fixed value of $k$, we additionally impose the following stability condition on the blended dynamics \eqref{eq:cont_emer}, as motivated by the continuous-time blended dynamics theory \cite{JKim,JGLEE_tool}.
    \begin{assumption} \label{assum:contractive}
		The blended dynamics \eqref{eq:cont_emer} is contractive in the sense of \cite{Slotine}; i.e., there exist a symmetric positive definite matrix $H\in\mathbb{R}^{m\times m}$ and a constant $\rho_{\mathsf{s}}>0$ such that
        \begin{align*} 
          (\mathsf{s}_1 - \mathsf{s}_2)^\top H^2 \left(f_{\mathsf{s}} (\mathsf{s}_1) - f_{\mathsf{s}} (\mathsf{s}_2)\right)  \le -\rho_{\mathsf{s}} \left\| H(\mathsf{s}_1 - \mathsf{s}_2) \right\|^2  
        \end{align*}
        for all $\mathsf{s}_1, \mathsf{s}_2 \in \mathbb{R}^m$. 
	\end{assumption}

    Note that Assumption~\ref{assum:contractive} is imposed only on the blended dynamics and does not require all individual dynamics to be stable. Nevertheless, it is sufficient for the following convergence result.

    \begin{theorem} \label{thm:uniform}
       Let Assumptions \ref{assum:graph}--\ref{assum:contractive} hold. Then, there exist constants $\beta_1^{\rm E},\beta_1^{\rm I},\beta_2^{\rm E},\beta_2^{\rm I}>0$ and $\beta_3^{\rm E}\ge0$ such that, for any $x[0] \in \mathbb{R}^{mN}$ and $\mathsf{s}(0) \in \mathbb{R}^m$, the solution $x_i[\cdot,k]$ of \eqref{eq:simplest} and the solution $\mathsf{s}(\cdot)$ of \eqref{eq:cont_emer} satisfy   
     \begin{subequations}\label{eq:thm_1}
        \begin{align}
        &\limsup_{T\to\infty} 
        \left\|x_i[n_{T,k},k]-p_i\mathsf{s}(T)\right\|
        \notag\\
        & \quad \le \frac{p_i\beta_1^{\rm E}}{k-\bar k}
        +\frac{\beta_2^{\rm E}}{k-\bar k-k^*}
        +\frac{\beta_3^{\rm E}}{(k-\bar k-k^*)^2},
        \label{eq:thm_1_E}\\
        &\limsup_{T\to\infty} 
        \left\|x_i[n_{T,k},k]-p_i\mathsf{s}(T)\right\|
        \le
        \frac{p_i\beta_1^{\rm I}}{k}
        +\frac{\beta_2^{\rm I}}{k-k^*},
        \label{eq:thm_1_I}
        \end{align}
    \end{subequations}
    for all $i\in\mathcal{N}$, with $k>\bar k+k^*$ in the explicit case
    and $k>k^*$ in the implicit case. Here, 
    \begin{align}
         \bar k &:= \frac{(L_{\mathsf{s}} \|H\| \|H^{-1}\|)^2}{2 \rho_{\mathsf{s}} },  \label{bar_k}   \\
     \label{coupling_gain_lower_bound}
    k^*   & :=
    \frac{2L \|A\|  + (2{L^2}/{\rho_{\mathsf{s} } }) \| A^2 \| \|H\| \| H^{-1}\|}{1-\lambda_{\mathcal{W}}},
    \end{align}
    where $L_{\mathsf{s}}$ is the Lipschitz constant of $f_{\mathsf{s}}$, $L$ is the maximum of the Lipschitz constants of $f_i$, and $\lambda_{\mathcal{W}}
    := 1-\frac{1-\mu_{\mathcal{W}}}{\kappa_{\mathcal{W}}}
    \in[|\lambda_2|,1)$,
    with $\mathsf P:=pq^\top$, $\mathsf T$ denoting any nonsingular matrix satisfying
    $\mu_{\mathcal{W}} := \|\mathsf T(\mathcal{W}-\mathsf P)\mathsf T^{-1}\|<1$, and
    $ \kappa_{\mathcal{W}} := \|\mathsf P\|
    \|\mathsf T(I_N-\mathsf P)\|
    \|\mathsf T^{-1}\|$.\footnote{Such a matrix $\mathsf T$ exists since $\rho(\mathcal{W}-\mathsf P)=|\lambda_2|<1$; its choice affects $k^*,\beta_2^{\rm E},\beta_3^{\rm E},\beta_2^{\rm I}$, but not $\bar k,\beta_1^{\rm E},\beta_1^{\rm I}$. For fixed $k$, one may take the infimum of the bound over such $\mathsf T$ satisfying the corresponding condition on $k$.}
    \end{theorem}

    Theorem~\ref{thm:uniform} shows that, for sufficiently large $k$, the approximation error between $x_i[n_{T,k},k]$ and $p_i\mathsf{s}(T)$ remains small for all sufficiently large $T$, with a bound that can be made arbitrarily small by increasing $k$. Its proof also establishes that the approximation is uniform in time.

    In both \eqref{eq:thm_1_E} and \eqref{eq:thm_1_I}, the first term corresponds to the explicit or implicit Euler discretization error between \eqref{eq:emer_dyn} and \eqref{eq:cont_emer}, while the remaining terms correspond to the synchronization error between \eqref{eq:simplest} and \eqref{eq:emer_dyn}, which is driven by the disagreement state $\tilde \xi$. Thus, $\bar k$ is associated with the explicit Euler discretization and $k^*$ is associated with synchronization. This gives the sufficient condition $k>\bar k+k^*$ in the explicit case. In the implicit case, $k>k^*$ also guarantees that \eqref{eq:simplest_implict} is well posed.\footnote{For fixed $x_i[n]$ and neighbor states, \eqref{eq:simplest_implict} is uniquely solvable because the map $I_m-\frac{a_i}{k}f_i(\cdot)$ is invertible when $L|a_i|<k$. Since $k^*\ge2L\|A\|$, this condition holds whenever $k>k^*$.} The following lemma shows that this Euler interpretation indeed holds. 

       \begin{lemma}\label{lem_error}
       Under Assumption \ref{assum:contractive}, for any $s[0],\mathsf{s}(0)\in\mathbb{R}^m$, the solution $s[\cdot,k]$ of \eqref{eq:emer_dyn} and the solution $\mathsf{s}(\cdot)$ of \eqref{eq:cont_emer} satisfy
        \begin{subequations}\label{eq:error_bounds}
        \begin{align}
        &\limsup_{T\to\infty} \left\|s\left[n_{T,k},k\right]-\mathsf{s}(T)\right\|
         \le \frac{\beta^{\mathrm E}_1}{k-\bar{k}},\quad \forall k>\bar{k}, \label{eq:error_bounds_explicit}\\
        &\limsup_{T\to\infty} \left\|s\left[n_{T,k},k\right]-\mathsf{s}(T)\right\|
         \le \frac{\beta^{\mathrm I}_1}{k},\quad \quad~ \forall k>0,
         \label{eq:error_bounds_implicit}
        \end{align}
        \end{subequations}
        where $\beta_1^{\mathrm E},\beta_1^{\mathrm I}>0$ are constants given in Theorem~\ref{thm:uniform}.
    \end{lemma}

    \begin{remark}
       The value of $k^*$ in \eqref{coupling_gain_lower_bound} well reflects the underlying intuition that $k^*$ must increase as the Lipschitz constant $L$ or $\|A\|$ becomes larger. It also must increase as the network connectivity becomes weaker, since then $|\lambda_2|$ approaches one and $\lambda_{\mathcal{W}} \in[|\lambda_2|,1)$ consequently approaches one. By contrast, $k^*$ can decrease when the contraction rate $\rho_{\mathsf{s}}$ becomes larger. 
    \end{remark}

    \begin{remark}
    The bounds in Theorem~\ref{thm:uniform} are invariant under the normalization of $p$ and $q$. Indeed, for any $\nu>0$, let $p'=\nu p$, $q'=q/\nu$, $s'[n]=s[n]/\nu$, and $\mathsf{s}'(t)=\mathsf{s}(t)/\nu$. Then $s'$ and $\mathsf{s}'$ satisfy \eqref{eq:emer_dyn} and \eqref{eq:cont_emer} under the new normalization, while $p_i's'[n]=p_is[n]$ and $p_i'\mathsf{s}'(t)=p_i\mathsf{s}(t)$. The constants $\beta_1^{\rm E},\beta_1^{\rm I}$ scale by $1/\nu$, whereas $\beta_2^{\rm E},\beta_3^{\rm E}, \beta_2^{\rm I},\bar k,k^*$, and $\lambda_{\mathcal{W}}$ remain unchanged. Hence the bounds in \eqref{eq:thm_1} are invariant.
    \end{remark}

\section{Decentralized Design of the Algorithm} \label{section:design}

Based on the analysis performed in the previous section, we now turn to the design of \eqref{eq:simplest}. We begin by specifying the desired continuous-time blended dynamics as
\begin{align}\label{eq:cont_dyn_emer}
\dot{\mathsf{s}}(t)
=  \sum_{i=1}^N w_i f_i(v_i\mathsf{s}(t)),
\end{align}
where $w_i>0$ and $v_i>0$ are the desired weights and scaling factors, respectively. 
In this setup, Section~\ref{subsection:decent_design} provides a decentralized method for choosing $a_i,b_i,c_i$ to realize \eqref{eq:cont_dyn_emer}. Section~\ref{subsection:special_case} presents three special cases used in Section~\ref{section:applications}. Finally, Section~\ref{subsection:trade-off} examines how $k$ affects the iteration count sufficient to keep the approximation error bound below a prescribed accuracy.

\subsection{Decentralized computation of $a_i$, $b_i$, $c_i$} \label{subsection:decent_design}

We first provide a necessary and sufficient condition under which the blended dynamics of \eqref{eq:simplest} coincides with \eqref{eq:cont_dyn_emer}.

\begin{proposition}\label{prop:decent_design}
Suppose that Assumption~\ref{assum:graph} holds. Given $w_i>0$ and $v_i>0$, $i\in\mathcal{N}$, and $\tau_{\mathsf{s}}>0$, the following statements are equivalent.
\begin{enumerate}
\item[(i)] Assumptions~\ref{assum:b,c} and~\ref{assum:W} hold, and \eqref{eq:simplest} has the blended dynamics \eqref{eq:cont_dyn_emer} for any choice of $f_i$, up to a fixed positive rescaling of $\mathsf{s}$.

\item[(ii)] There exist $\delta_i$ satisfying
\begin{align} \label{eq:delta}
0<\delta_i\le {1}/{\tilde d_i},\qquad i\in\mathcal{N},
\end{align}
with strict inequality for at least one $i\in\mathcal{N}$, such that
\begin{align} \label{eq:abc}
a_i=\frac{ \tau_{\mathsf{s}}w_i v_i\delta_i}{\tilde\zeta_i}
 \sum_{j=1}^N 
\frac{\tilde\zeta_j}{\delta_j}, \quad
b_i=1-\delta_i\tilde d_i, \quad
c_i=\delta_i.
\end{align}
\end{enumerate}
Here, $\tilde d_i
    :=
    \frac{1}{v_i}\sum_{j=1}^{N}\alpha_{ij}v_j$, $i\in\mathcal{N}$, and $\tilde\zeta=\operatorname{col}(\tilde\zeta_i)_{i\in\mathcal{N}}$ is the normalized left eigenvector of $\tilde{\mathcal{L}}$ associated with the zero eigenvalue, satisfying
    $\tilde\zeta^\top\tilde{\mathcal{L}}=0_N^\top$ and $1_N^\top\tilde\zeta=1$, where $\tilde{\mathcal{L}}:=\tilde{\mathcal{D}}-\tilde{\mathcal{A}}$, $\tilde{\mathcal{D}}:=\operatorname{diag}(\tilde d_i)_{i\in\mathcal{N}}$, $\tilde{\mathcal{A}}:=V^{-1}\mathcal{A} V$, and $V:=\operatorname{diag}(v_i)_{i\in\mathcal{N}}$.
\end{proposition}

   Following \eqref{eq:abc}, to obtain a locally implementable update, let us choose the auxiliary time-scale parameter as
    \begin{align}\label{eq:design_tau}
    \tau_{\mathsf{s}} :=  \frac{1}{\sum_{i=1}^{N}\tilde\zeta_i/\delta_i}.
   \end{align}
    With this choice, we propose the distributed algorithm as
    \begin{align} \label{eq:decent_form_implicit}
    \begin{split}
        x_i[n+1]
        ={}&
        \frac{w_i v_i\delta_i}{k\tilde\zeta_i}
        f_i(x_i[\,\cdot\,])
        +(1-\delta_i\tilde d_i)x_i[n] \\
        &+
        \delta_i \sum_{j\in\mathcal{N}_i}\alpha_{ij}x_j[n],
    \end{split}
    \end{align}
    for each $i\in \mathcal{N}$. Although the choice of $\tau_{\mathsf{s}}$ in \eqref{eq:design_tau} depends on global information, it is only an auxiliary parameter and cancels from \eqref{eq:decent_form_implicit}; hence, it is not needed for implementation. 
    The following decentralized procedure determines $a_i,b_i,c_i$ for a common $k$ fixed before execution. Selecting $k$ to meet the sufficient conditions of Theorem~\ref{thm:uniform} involves the global quantities in $\bar k$ and $k^*$ and is not part of this procedure.
   \newcounter{implementationstep}
    \begin{list}{Step~\arabic{implementationstep})}{
      \usecounter{implementationstep}
      \setlength{\labelwidth}{3.5em}
      \setlength{\labelsep}{0.5em}
      \setlength{\leftmargin}{4em}
      \setlength{\itemsep}{0.5em}
      \setlength{\parsep}{0pt}
      \setlength{\topsep}{0.5em}
    }%
    \item\label{step:1} 
    Each agent~$i$ is assigned $w_i$ and $v_i$. It exchanges $v_i$ once with its neighbors and computes $\tilde d_i = (1/v_i)\sum_{j\in\mathcal{N}_i}\alpha_{ij}v_j$.
    
    \item\label{step:2}
    Each agent~$i$ chooses $\delta_i$ such that $0<\delta_i< 1/\tilde d_i$.

    \item\label{step:3}
    Each agent~$i$ estimates $1/\tilde\zeta_i$ using the distributed algorithm proposed in Section~\ref{subsection:zeta}.
    
    \item\label{step:4}
    Each agent~$i$ implements \eqref{eq:decent_form_implicit} with its own state and the received neighbor states.
    \end{list}

    With the exact values of $1/\tilde{\zeta}_i$, the convergence results in Section~\ref{section:2} apply to \eqref{eq:decent_form_implicit}. Indeed, by Proposition~\ref{prop:decent_design}, \eqref{eq:decent_form_implicit} has \eqref{eq:cont_dyn_emer} as its blended dynamics. In particular, if \eqref{eq:cont_dyn_emer} satisfies Assumption~\ref{assum:contractive}, then Theorem~\ref{thm:uniform} guarantees that, for sufficiently large $k$, $x_i[n]$ eventually stays close to $v_i\mathsf{s}(t_{n,k})$, where $t_{n,k}:=nh_k =n \tau_{\mathsf{s}}/k$, and $\tau_{\mathsf{s}}$ is given in \eqref{eq:design_tau}.

\begin{proof}[Proof of Proposition~\ref{prop:decent_design}]
\noindent{((i) $\Rightarrow$ (ii)).} Suppose that (i) holds. Choose the rescaling in (i) so that \eqref{eq:cont_emer} and \eqref{eq:cont_dyn_emer} coincide exactly. Since this holds for arbitrary $f_i$, their arguments and coefficients must agree, giving $p_i=v_i$ and $q_i a_i/\tau_{\mathsf{s}}=w_i$ for all $i\in\mathcal{N}$. Hence, $p=v:=\operatorname{col}(v_i)_{i \in \mathcal{N}}$ and $\mathcal{W} v=v$. Since $\mathcal{W}=B+C\mathcal{A}$, this gives
$b_i v_i + c_i\sum_{j=1}^N\alpha_{ij}v_j=v_i$, $i\in\mathcal{N}$,
so that $b_i=1-c_i\tilde d_i$. Letting $\delta_i:=c_i$, Assumption~\ref{assum:b,c} gives \eqref{eq:delta}, including strict inequality for at least one $i\in\mathcal{N}$.

With $\Delta:=\operatorname{diag}(\delta_i)_{i\in\mathcal{N}}$, the relations $b_i=1-\delta_i\tilde d_i$ and $c_i=\delta_i$ give
$\mathcal{W}= V(I_N-\Delta\tilde{\mathcal{L}})V^{-1}$ and 
$q= \frac{V^{-1}\Delta^{-1}\tilde\zeta}{1_N^\top\Delta^{-1}\tilde\zeta}$,
where the expression for $q$ follows from $\tilde\zeta^\top\tilde{\mathcal{L}}=0_N^\top$ and $q^\top v=1$. Hence, $q_i = \frac{\tilde\zeta_i}{v_i\delta_i}/ \sum_{j=1}^N(\tilde\zeta_j/\delta_j)$. Combining this with $q_i a_i/ \tau_{\mathsf{s}} =w_i$ gives \eqref{eq:abc}. Thus, (ii) holds.

\noindent{((ii) $\Rightarrow$ (i)).}
Suppose that (ii) holds. From \eqref{eq:delta} and \eqref{eq:abc}, we have $c_i>0$, $b_i\ge0$ for all $i\in\mathcal{N}$, and $b_i>0$ for at least one $i\in\mathcal{N}$. Hence, Assumption~\ref{assum:b,c} holds. Moreover, 
$(\mathcal{W} v)_i = (1-\delta_i\tilde d_i)v_i + \delta_i\sum_{j=1}^N\alpha_{ij}v_j = v_i$, so $\mathcal{W} v=v$. Since $\mathcal{W}$ is primitive and $v\in \mathbb{R}^N_{>0}$ is a positive eigenvector of $\mathcal{W}$ associated with the eigenvalue one, Proposition~\ref{lem_Perron} implies that $\rho(\mathcal{W})=1$. Hence, Assumption~\ref{assum:W} holds, and we may take $p=v$.

Moreover, from $\mathcal{W}=V(I_N-\Delta\tilde{\mathcal{L}})V^{-1}$ and $q^\top v =1$, we obtain $q_i = {\tilde\zeta_i}/{(v_i\delta_i \sum_{j=1}^N(\tilde\zeta_j/\delta_j))}$. Together with \eqref{eq:abc}, this gives $q_i a_i/\tau_{\mathsf{s}}=w_i$ for all $i\in\mathcal{N}$. Therefore, \eqref{eq:cont_emer} coincides with \eqref{eq:cont_dyn_emer}, and (i) follows.
\end{proof}

\subsection{Special cases} \label{subsection:special_case}

The general design in \eqref{eq:decent_form_implicit} takes simpler forms in three special cases, summarized in Table~\ref{tab:special_cases}. Here, $\delta_i$ is selected locally as described in the preceding subsection.

\begin{table}[t]
\centering
\caption{Design parameter choices in the special cases}
\label{tab:special_cases}
\setlength{\tabcolsep}{4pt}
\begin{tabular}{|c|c|c|}
\hline
Setting
&
$(\tilde d_i,\tilde\zeta_i)$
&
$(a_i,b_i,c_i)$
\\
\hline \hline
Averaged, undirected
&
$(d_i,1/N)$
&
$(\delta_i,1-\delta_i d_i,\delta_i)$
\\
Weighted, directed
&
$(d_i,\zeta_i)$
&
$(w_i\delta_i/\zeta_i,1-\delta_i d_i,\delta_i)$
\\
Scaled, directed
&
$(d_i^{\text{out}},p_i)$
&
$(\delta_i,1-\delta_i d_i^{\text{out}},\delta_i)$
\\
\hline
\end{tabular}
\end{table}

\subsubsection{Averaged blended dynamics over undirected graphs} \label{subsection:special_1}

For an undirected and connected graph, consider the desired averaged dynamics
\begin{align}\label{eq:avg_blended}
    \dot{\mathsf{s}}(t)
    =
    \frac{1}{N}
     \sum_{i=1}^{N}f_i(\mathsf{s}(t)).
\end{align}
This corresponds to $w_i={1}/{N}$ and $v_i=1$ for all $i\in\mathcal{N}$. Consequently, we have $\tilde d_i=d_i$ and $\tilde\zeta_i=1/N$. Substituting these quantities into \eqref{eq:decent_form_implicit} gives the update
    \begin{equation}
    \begin{aligned}\label{eq:avg_algorithm}
    x_i[n+1]
    ={}&
    \frac{\delta_i}{k}f_i(x_i[\cdot])
    +
    (1-\delta_i d_i)x_i[n]
    \\
    &+
    \delta_i
     \sum_{j\in\mathcal{N}_i}\alpha_{ij}x_j[n], \qquad i \in \mathcal{N}.
    \end{aligned}
    \end{equation}
This case is particularly simple because $w_i$, $v_i$, and $\tilde\zeta_i$ all cancel out. Hence Steps~\ref{step:1} and \ref{step:3} of the procedure above are not required: the agents need neither exchange $v_i$ nor estimate $1/\tilde\zeta_i$. Each agent only needs its local degree $d_i$ to choose $\delta_i$ according to Step~\ref{step:2}. This simplified structure will be used for network size estimation in Section~\ref{subsection:netsize_est}.

    \begin{remark}
    For an undirected and connected graph, the averaged dynamics \eqref{eq:avg_blended} also arises as the blended dynamics of the continuous-time multi-agent system studied in \cite{JKim,JGLEE_tool}:
    \[
    \dot x_i(t)
    =
    f_i(x_i(t))
    +
    k \sum_{j\in\mathcal{N}_i}
    \alpha_{ij}\bigl(x_j(t)-x_i(t)\bigr), \quad i\in \mathcal{N},
  \]
  where $k>0$ is the coupling gain. This continuous-time system is related to \eqref{eq:avg_algorithm}. In particular, its forward Euler discretization with sampling time $h=\delta_{\mathsf c}/k$\footnote{With a fixed sampling time, the linear coupling part of the Euler discretization may become unstable as $k$ increases. The $k$-dependent choice $h=\delta_{\mathsf c}/k$ keeps $hk=\delta_{\mathsf c}$ fixed, so that its stability can be maintained independently of $k$.} coincides with the explicit form of \eqref{eq:avg_algorithm} for $\delta_i=\delta_{\mathsf c}$, $i\in\mathcal{N}$. The condition $0<\delta_{\mathsf c}<\min_{i\in\mathcal{N}}1/d_i$ ensures stability of the linear coupling dynamics. In contrast, \eqref{eq:avg_algorithm} allows each agent to choose $\delta_i$ locally using only $d_i$. Both choices yield \eqref{eq:avg_blended}, but have different auxiliary sampling times $h_k$.
    \end{remark}

\subsubsection{Weighted blended dynamics over strongly connected digraphs} \label{subsection:special_2}

For a strongly connected digraph, consider the desired weighted blended dynamics
\begin{align}\label{eq:weighted_blended}
    \dot{\mathsf{s}}(t)
    =  \sum_{i=1}^{N}w_i f_i(\mathsf{s}(t)),
\end{align}
where $w_i>0$. This corresponds to $v_i=1$ for all $i\in\mathcal{N}$, and hence we obtain $\tilde d_i=d_i$ and $\tilde\zeta_i=\zeta_i$, where $\zeta=\operatorname{col}(\zeta_i)_{i\in\mathcal{N}}$ is the normalized left eigenvector of $\mathcal{L}$ associated with the zero eigenvalue. Substituting these quantities into \eqref{eq:decent_form_implicit} gives the update 
    \begin{equation}
    \begin{aligned}\label{eq:weighted_algorithm}
    x_i[n+1]
    ={}&
    \frac{w_i\delta_i}{k\zeta_i}f_i(x_i[\cdot])
    +
    (1-\delta_i d_i)x_i[n]
    \\
    & + \delta_i \sum_{j\in\mathcal{N}_i}\alpha_{ij}x_j[n], \qquad i \in \mathcal{N}.
    \end{aligned}
    \end{equation}
The additional quantity $1/\zeta_i$, which compensates for the graph asymmetry, can be obtained using the distributed algorithm proposed in Section~\ref{subsection:zeta}. That estimator is
itself obtained from \eqref{eq:weighted_algorithm}: choosing
$w_i=\zeta_i$ makes the coefficient $w_i/\zeta_i$ equal to one,
so $\zeta_i$ is not needed for its implementation. The form \eqref{eq:weighted_algorithm} will be used in Section~\ref{subsection:DWLS} to construct a distributed weighted least-squares solver.

\subsubsection{Scaled blended dynamics over strongly connected digraphs}
\label{subsection:special_3}

    This case is motivated by PageRank computation. In this application, $p_i$ is the normalized value of $r_i/d_i^{\text{out}}$, where $r_i$ denotes the PageRank score and $d_i^{\text{out}}$ is the out-degree of agent~$i$. Its precise definition is given in Section~\ref{subsection:PageRank}. For a strongly connected digraph, $p=\operatorname{col}(p_i)_{i\in\mathcal{N}}\in\mathbb{R}_{>0}^{N}$ satisfies
    $\mathcal{L}_{\text{out}}p=0_N$ and $1_N^\top p=1$,
   where $\mathcal{L}_{\text{out}}:=\mathcal{D}_{\text{out}}-\mathcal{A}$ and $\mathcal{D}_{\text{out}}:=\operatorname{diag}(d_i^{\text{out}})_{i\in\mathcal{N}}$. Now, consider the desired scaled blended dynamics
    \begin{align}\label{eq:outdegree_blended}
        \dot{\mathsf{s}}(t)
        =
         \sum_{i=1}^{N}f_i(p_i\mathsf{s}(t)).
    \end{align}
    This corresponds to $w_i=1$ and $v_i=p_i$ for all $i\in\mathcal{N}$. Letting $P:=\operatorname{diag}(p_i)_{i\in\mathcal{N}}$, the relation
    $\mathcal{L}_{\text{out}}p=0_N$ yields
    $\tilde d_i=d_i^{\text{out}}$, $\tilde{\mathcal{L}}=P^{-1}\mathcal{L}_{\text{out}}P$,
    and $\tilde\zeta_i=p_i$.
    Substituting these quantities into \eqref{eq:decent_form_implicit} gives the update
    \begin{equation}
    \begin{aligned}\label{eq:outdegree_algorithm}
    x_i[n+1]
    ={}&
    \frac{\delta_i}{k}f_i(x_i[\cdot])
    +
    (1-\delta_i d_i^{\text{out}})x_i[n]
    \\
    & + \delta_i \sum_{j\in\mathcal{N}_i}\alpha_{ij}x_j[n].
    \end{aligned}
    \end{equation}   
    Although $p_i$ determines the desired scaling in \eqref{eq:outdegree_blended}, it cancels out in \eqref{eq:outdegree_algorithm} and therefore need not be estimated for the purpose of implementation. Thus, each agent only utilizes its local out-degree $d_i^{\text{out}}$. Section~\ref{subsection:PageRank} uses this property to obtain the scaled PageRank score and then recover the PageRank score through a local rescaling.    

    \subsection{Optimization of the transient behavior} \label{subsection:trade-off}
    
    The approximation error bounds in Theorem~\ref{thm:uniform} decrease as $k$ increases, which may suggest taking $k$ as large as possible. However, the sampling time $h_k=\tau_{\mathsf{s}}/k$ also decreases, so a fixed time interval of the blended dynamics corresponds to more iterations. Thus, a larger $k$ does not necessarily reduce the minimum number of iterations required to guarantee a given accuracy for all subsequent iterations.

     To study this trade-off, we use the finite-iteration error bound $e(n,k,x[0],s[0],\mathsf{s}(0))$ given in
    \eqref{eq:app-error-bound} of Appendix~\ref{app:proof-theorem2}. It combines bounds on the synchronization error between $x_i[n,k]$ and $p_i s[n,k]$ and the discretization error between
    $p_i s[n,k]$ and $p_i\mathsf{s}(t_{n,k})$, where $t_{n,k}:=nh_k$, so that
    \[
    \max_{i\in\mathcal{N}} \|x_i[n,k]-p_i\mathsf{s}(t_{n,k})\| \le e(n,k,x[0],s[0],\mathsf{s}(0)).
    \]
    This bound consists of the common asymptotic error bound in Theorem~\ref{thm:uniform} and nonnegative transient terms depending on the initial conditions.

    For compact sets $\mathcal K_x\subset\mathbb{R}^{mN}$ and $\mathcal K_{\mathsf{s}}\subset\mathbb{R}^m$, define the worst-case error bound as
    \begin{align}
    e_{\mathcal K}(n,k) := \sup_{\substack{x[0]\in\mathcal K_x\\\mathsf{s}(0)\in\mathcal K_{\mathsf{s}}}} \inf_{s[0]\in\mathbb{R}^m} e(n,k,x[0],s[0],\mathsf{s}(0)),    
    \label{eq:eK}
    \end{align}
    where the supremum is taken over the prescribed initial sets, while the infimum is taken over the initial value $s[0]$ of the auxiliary dynamics \eqref{eq:emer_dyn}.

    For each fixed $k$ in the corresponding range, the worst-case bound $e_{\mathcal K}(n,k)$ converges as $n\to\infty$ to the common asymptotic error bound of Theorem~\ref{thm:uniform}, denoted by $e_\infty(k)$. For the explicit and implicit updates, respectively,
    \begin{subequations}
    \label{eq:e-infty}
    \begin{align}
    e_\infty(k)
    & := \frac{p_{\max}\beta_1^{\rm E}}{k-\bar k} +\frac{\beta_2^{\rm E}}{k-\bar k-k^*} + \frac{\beta_3^{\rm E}}{(k-\bar k-k^*)^2}, 
    \label{eq:e-infty-E}\\
    e_\infty(k)
    & := \frac{p_{\max}\beta_1^{\rm I}}{k} +\frac{\beta_2^{\rm I}}{k-k^*},
    \label{eq:e-infty-I}
    \end{align}
    \end{subequations}
    where $p_{\max}:=\max_{i\in\mathcal{N}}p_i$, and the respective ranges are $k>\bar k+k^*$ and $k>k^*$. In either case, $e_\infty(\cdot)$ is continuous and strictly decreasing from infinity to zero over its domain. Thus, each $\epsilon>0$ determines a unique $k_\epsilon$ satisfying $e_\infty(k_\epsilon)=\epsilon$, and $e_\infty(k)<\epsilon$ holds exactly when $k>k_\epsilon$.

    Although $e_\infty(k)$ decreases with $k$, the transient part $e_{\mathcal K}(n,k)-e_\infty(k)$ need not decrease with $k$ for a fixed $n$.
    Now, for $k>k_\epsilon$, let $\mathbf N_\epsilon(k)$ denote the smallest number of iterations after which the bound remains at most $\epsilon$:
    \begin{equation}
    \label{eq:N_epsilon}
    \mathbf N_\epsilon(k)
    :=
    \min\left\{
    n_0\in\mathbb Z_{\ge0}:
    e_{\mathcal K}(n,k)\le\epsilon
    \text{ for all }n\ge n_0
    \right\}.
    \end{equation}
    By the preceding discussion, $\mathbf N_\epsilon(k)$ is finite for every $k>k_\epsilon$. The following theorem further analyzes its behavior with respect to $k$ in the two extreme cases.

    \begin{theorem}
    \label{thm:iteration}
    Let Assumptions~\ref{assum:graph}--\ref{assum:contractive} hold, and let $\mathcal K_x\subset\mathbb{R}^{mN}$ and $\mathcal K_{\mathsf{s}}\subset\mathbb{R}^m$ be compact. For either update and every $\epsilon>0$, let $e_\infty(\cdot)$ be defined by \eqref{eq:e-infty} and let $k_\epsilon$ be the unique value satisfying $e_\infty(k_\epsilon)=\epsilon$. Then, $\mathbf N_\epsilon(k)$ in \eqref{eq:N_epsilon} is finite for every $k>k_\epsilon$. Moreover, the following holds under the conditions stated below:
    \[
    \lim_{k\to\infty}
    \mathbf N_\epsilon(k)
    =
    \infty, \qquad
    \lim_{k\searrow k_\epsilon}
    \mathbf N_\epsilon(k)=\infty.
    \]
    The first limit holds if $p_{\max}\|H^{-1}\|\allowbreak \sup_{x[0]\in\mathcal K_x,\,\mathsf{s}(0)\in\mathcal K_{\mathsf{s}}} \allowbreak\|H(\xi_1[0]-\mathsf{s}(0))\|>\epsilon$, where $\xi_1[0]=q^\top_{\otimes m} x[0]$. The second limit holds if $\sup_{\mathsf{s}(0)\in\mathcal K_{\mathsf{s}}}\allowbreak \|H\mathsf{s}(0)\|>\|Hf_{\mathsf{s}}(0)\|/\rho_{\mathsf{s}}$.
    \end{theorem}

    The first condition concerns the difference between the initial weighted sum $\xi_1[0]$ of the agent states and the initial value $\mathsf{s}(0)$ of the blended dynamics \eqref{eq:cont_emer}, rather than disagreement among individual agents. The increment of $\xi_1$ in \eqref{eq:xi_1_dynamics} is scaled by $1/k$, and $\mathsf{s}$ is evaluated at $t_{n,k}=n\tau_{\mathsf{s}}/k$. Thus, for fixed $n$, both $\xi_1[n]$ and $\mathsf{s}(t_{n,k})$ stay near their initial values as $k\to\infty$. Under the first condition, for each fixed $n$, the worst-case bound $e_{\mathcal K}(n,k)$ remains above $\epsilon$ for all sufficiently large $k$, so the guaranteed iteration count $\mathbf N_\epsilon(k)$ diverges as $k\to\infty$. This initial difference vanishes under the initialization in Theorem~\ref{thm:lim_sol}.

    When both conditions in Theorem~\ref{thm:iteration} hold, $\mathbf N_\epsilon(k)$ diverges at both ends of $(k_\epsilon,\infty)$ and therefore attains its minimum at a finite $k>k_\epsilon$. Although this analysis is based on a possibly conservative error bound derived in Appendix~\ref{app:proof-theorem2}, the same tendency is also demonstrated in Example~\ref{ex:size_estimation_two_agent} in Section~\ref{subsection:netsize_est} using a directly derived bound.

    \section{Applications} \label{section:applications}

    We now apply the decentralized design methodology proposed in Section~\ref{subsection:special_case} to four distributed problems. For each problem, we first choose a blended dynamics whose equilibrium gives the desired quantity, and then derive a corresponding implicit update in closed form that can be implemented using only local information.
    
\subsection{Network size estimation} \label{subsection:netsize_est}

The network size~$N$ is required in many distributed computations, but it is a global quantity that is not locally available to individual agents. For an undirected connected graph, consider the blended dynamics
\begin{equation}\label{eq:size_blended}
    \dot{\mathsf{s}}(t)
    =
    -\frac{1}{N}\mathsf{s}(t)+1,
\end{equation}
whose unique equilibrium is $N$.

Following the averaged design specialized in Section~\ref{subsection:special_1}, choose $f_1(x_1)=-x_1+1$ and $f_i(x_i)=1$ for $i=2,\ldots,N$. Then,
    $\frac{1}{N}\sum_{i=1}^N f_i(\mathsf{s})
    =\frac{1}{N}\left( (-\mathsf{s}+1) + (N-1) \right)=
    -\frac{1}{N}\mathsf{s}+1$, giving \eqref{eq:size_blended}. With these choices, the implicit form of \eqref{eq:avg_algorithm} for agent~$1$ can be explicitly written as
    \begin{equation}
     \begin{aligned}
        x_1[n+1]
        =\frac{k}{k+{\delta_1}}
        \bigg(
         & \frac{\delta_1}{k}
         +(1-\delta_1 d_1)x_1[n] \\
         & ~ +\delta_1 \sum_{j\in\mathcal N_1}\alpha_{1j}x_j[n]
        \bigg),
        \end{aligned}
        \label{eq:size_agent1}
    \end{equation}
    while agents $i=2,\ldots,N$ run
    \begin{equation}\label{eq:size_agenti}
    x_i[n+1]
    =
    \frac{\delta_i}{k}
    +(1-\delta_i d_i)x_i[n]
    +\delta_i
     \sum_{j\in\mathcal N_i}
    \alpha_{ij}x_j[n].
    \end{equation}
    Here, $\delta_i>0$ is chosen as in Proposition~\ref{prop:decent_design} with $\tilde d_i=d_i$. 
    
    Theorem~\ref{thm:uniform} guarantees that each state eventually remains close to $N$ for sufficiently large $k$. Once the approximation error is smaller than $0.5$, each agent recovers $N$ exactly by rounding $x_i[n]$ to the nearest integer. Related continuous-time and discrete-time multi-step estimators based on the same blended dynamics were studied in~\cite{DGLee,JWKim}.

    Theorem~\ref{thm:iteration} shows that the sufficient iteration count does not necessarily decrease as $k$ increases. The following two-agent example shows the trade-off discussed in Theorem~\ref{thm:iteration} using a directly derived error bound.

\begin{example}\label{ex:size_estimation_two_agent}

Consider the network-size estimator \eqref{eq:size_agent1}--\eqref{eq:size_agenti} on a two-agent graph with $\alpha_{12}=\alpha_{21}=1$. Let $\delta_1=\delta_2=0.5$, and suppose that $x_i[0]\in[0.5,\bar N+0.5]$ for $i=1,2$, where $\bar N\ge 2$. Then, for each $k>2$,
$\left\lceil x_i[n]\right\rfloor=2$ for $i=1,2$ and all $n\geq \mathbf{N}_{0.5}(k)$.
Moreover,
\[
\lim_{k\to\infty}\mathbf{N}_{0.5}(k)=\infty, \qquad
\lim_{k\searrow2}\mathbf{N}_{0.5}(k)=\infty,
\]
where $\mathbf N_{0.5}(k):=\left\lfloor\Gamma_{0.5}(k)\right\rfloor+1$ and
\[
\Gamma_{0.5}(k)
:=
1+
\ln\left(\frac{2k\bar N}{k-2}\right)
\left[
\ln\left(\frac{4k+2}{4k+1}\right)
\right]^{-1}.
\]

\end{example}

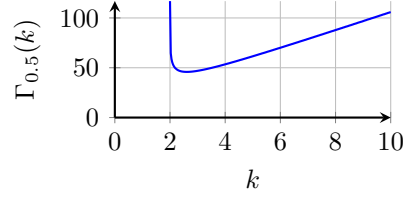
\begin{figure}[!t]
\centering
\begin{tikzpicture}
\begin{axis}[
    width=0.62\linewidth,
    height=0.37\linewidth,
    xmin=0, xmax=10,
    ymin=0,
    xlabel={$k$},
    ylabel={$\Gamma_{0.5}(k)$},
    domain=2.0001:10,
    samples=300,
    thick,
    grid=both,
    axis lines=left,
    enlargelimits=false,
]
\addplot[
    blue,
]
{1 + ln(2*x*5/(x-2)) / ln((4*x+2)/(4*x+1))};
\end{axis}
\end{tikzpicture}
\caption{The bound $\Gamma_{0.5}(k)$ versus $k$ ($\bar N = 5$).} 
\label{fig:N05_vs_k}
\end{figure}

Here, $\mathbf N_{0.5}(k)$ denotes a sufficient iteration count derived directly in Appendix~\ref{ex:size_est_two_proof}. The two limits imply that it is minimized at a finite $k>2$. Figure~\ref{fig:N05_vs_k} plots the continuous threshold $\Gamma_{0.5}(k)$ for $\bar N=5$.

\subsection{Estimation of $1/\zeta_i$} \label{subsection:zeta}

The directed-graph design in Section~\ref{subsection:special_2} requires agent~$i$ to know $1/\zeta_i$, although $\zeta_i$ depends on the entire graph and is therefore global information. To estimate these values in a decentralized manner, let us first consider the blended dynamics
\begin{equation}\label{eq:zeta_inverse_blended}
    \dot{\mathsf{s}}(t)
    =
    -D_\zeta\mathsf{s}(t)+1_N,
\end{equation}
where $D_\zeta:=\operatorname{diag}(\zeta_i)_{i\in\mathcal{N}}$. Its unique equilibrium is $D_\zeta^{-1}1_N=\operatorname{col}(1/\zeta_i)_{i\in\mathcal{N}}$.

Assume that $N$ is known and that each agent has a distinct identifier. Let $\mathbf{e}_i\in\mathbb{R}^N$ denote the $i$-th standard basis vector and define $\mathbf{E}_i:=\mathbf{e}_i\mathbf{e}_i^\top$. In the design in Section~\ref{subsection:special_2}, choose $f_i(x_i)=-\mathbf{E}_i x_i +1_N$ and $w_i=\zeta_i$.
Then, 
$\sum_{i=1}^{N}\zeta_i f_i(\mathsf{s})
=
-D_\zeta\mathsf{s}+1_N,$
which gives \eqref{eq:zeta_inverse_blended}. Since $w_i/\zeta_i=1$, the unknown value $\zeta_i$ cancels out when implementing the distributed algorithm via \eqref{eq:weighted_algorithm}. The resulting implicit form of \eqref{eq:weighted_algorithm} can be written explicitly as
\begin{equation} \label{eq:zeta_inverse_estimator}
\begin{aligned}
x_i[n+1]
= \left(I_N+\frac{\delta_i}{k}\mathbf{E}_i\right)^{-1} 
\bigg( & \frac{\delta_i}{k}1_N + (1-\delta_i d_i)x_i[n] \\
& +\delta_i \sum_{j\in\mathcal N_i}\alpha_{ij}x_j[n]
\bigg),
\end{aligned}
\end{equation}
where $i \in \mathcal{N}$ and $\delta_i>0$ is chosen as in Proposition~\ref{prop:decent_design} with $\tilde d_i=d_i$. 

Theorem~\ref{thm:uniform} guarantees that, for sufficiently large $k$, each $x_i[n]$ approximates the equilibrium of \eqref{eq:zeta_inverse_blended}. Since the only information needed by agent~$i$ is $1/\zeta_i$, the following stronger result holds.

\begin{proposition}\label{prop:zeta_inverse}
Suppose that Assumption~\ref{assum:graph} holds and that $\delta_i$ is chosen as in Proposition~\ref{prop:decent_design} with $\tilde d_i=d_i$. Then, for every $k>0$, the solution $x_i[\cdot]$ of \eqref{eq:zeta_inverse_estimator} satisfies
\[
\lim_{n\to\infty}\mathbf{e}_i^\top x_i[n]
= {1}/{\zeta_i},
\qquad
i\in\mathcal{N}.
\]
If, in addition, $x_i[0]\in\mathbb R_{>0}^N$ for every $i\in\mathcal{N}$, then $x_i[n]\in\mathbb R_{>0}^N$ for all $n\ge0$ and $i\in\mathcal{N}$.

\end{proposition}

The proof is given in Appendix~\ref{prop:zeta_inverse_proof}. Thus, agent~$i$ can use $\mathbf{e}_i^\top x_i[n]$ as its estimate of $1/\zeta_i$ for any $k>0$, without requiring global information on a lower bound for $k$. 

To implement Step~\ref{step:3} of Section~\ref{subsection:decent_design}, apply \eqref{eq:zeta_inverse_estimator} with $\alpha_{ij}$ and $d_i$ replaced by $\tilde{\alpha}_{ij}:= \alpha_{ij} v_j/v_i$ and $\tilde{d}_i := \sum_{j \in \mathcal{N}_i} \tilde{\alpha}_{ij}$, respectively.

\begin{remark}
The same construction can estimate the left eigenvector $\zeta$ rather than its reciprocal. Choosing $f_i(\mathsf{s})=-\mathsf{s}+\mathbf{e}_i$ and $w_i=\zeta_i$ gives the blended dynamics $\dot{\mathsf{s}}(t) = \sum_{i=1}^N \zeta_i(-\mathsf{s} (t)+ \mathbf{e}_i)= -\mathsf{s} (t) + \zeta$, whose unique equilibrium is $\zeta$. Hence, for sufficiently large $k$, each agent state approximates $\zeta$. Unlike the method in~\cite{APriolo}, our construction does not require specific initialization.
\end{remark}

\subsection{Weighted least-squares solver} \label{subsection:DWLS}

The proposed framework can also be used to construct distributed optimization algorithms from the gradient flow of a global objective. As an example, consider the weighted least-squares (WLS) problem
\begin{equation}
\begin{aligned}
\min_{x\in\mathbb{R}^m}
\frac{1}{2}\sum_{i=1}^N w_i |\mathbf{A}_i x-\mathbf{b}_i|^2,
\end{aligned}
\label{eq:wls_problem}
\end{equation}
where each agent~$i$ knows only $\mathbf{A}_i\in\mathbb{R}^{1\times m}$, $\mathbf{b}_i\in\mathbb{R}$, and $w_i>0$. Let $\mathbf{A}:=\operatorname{col}(\mathbf{A}_i)_{i\in\mathcal{N}}$,  $W:=\operatorname{diag}(w_i)_{i\in\mathcal{N}}$ and $\mathbf{b}:=\operatorname{col}(\mathbf{b}_i)_{i\in\mathcal{N}}$, and suppose that $\mathbf{A}$ has full column rank. 

The corresponding gradient flow is
\begin{equation}
\begin{aligned}
\dot{\mathsf{s}}(t)
&=
-\sum w_i \mathbf{A}_i^\top(\mathbf{A}_i \mathsf{s}(t)-\mathbf{b}_i) =
-\mathbf{A}^\top W(\mathbf{A} \mathsf{s}(t)-\mathbf{b}).
\end{aligned}
\label{eq:wls_gradient_flow}
\end{equation}
Since $\mathbf{A}^\top W \mathbf{A}$ is positive definite, \eqref{eq:wls_gradient_flow} has the unique equilibrium $x^*=(\mathbf{A}^\top W\mathbf{A})^{-1}\mathbf{A}^\top W\mathbf{b}$.

To obtain a distributed algorithm, apply the implicit form of \eqref{eq:weighted_algorithm} with $f_i(x)=-\mathbf{A}_i^\top(\mathbf{A}_i x-\mathbf{b}_i)$. With this choice, the resulting algorithm can be written explicitly as
\begin{equation}
\begin{aligned}
x_i[n+1]
 ={}&
\left(
I_m+\frac{w_i\delta_i}{k\zeta_i}
\mathbf{A}_i^\top\mathbf{A}_i
\right)^{-1}
\bigg(
\frac{w_i\delta_i}{k\zeta_i}
\mathbf{A}_i^\top\mathbf{b}_i
\\
& +(1-\delta_i d_i)x_i[n]
+\delta_i\sum_{j\in\mathcal N_i}
\alpha_{ij}x_j[n]
\bigg),
\end{aligned}
\label{eq:wls_distributed}
\end{equation}
where $i \in \mathcal{N}$ and $\delta_i>0$ is chosen as in Proposition~\ref{prop:decent_design} with $\tilde d_i=d_i$. The inverse is well defined, and Section~\ref{subsection:zeta} provides a decentralized estimator of $1/\zeta_i$.

With the exact value of $1/\zeta_i$, Theorem~\ref{thm:uniform} guarantees that, for sufficiently large $k$, each $x_i[n]$ eventually remains close to $x^*$. Thus, this example illustrates how the gradient flow of a weighted least-squares objective can be realized within the proposed framework over a strongly connected digraph.

\begin{remark}
 If the rank of matrix $\mathbf{A}$ is less than $m$, then the WLS problem \eqref{eq:wls_problem} has non-unique solutions, and \eqref{eq:wls_gradient_flow} no longer has a unique equilibrium. This case is not covered by the present analysis. 
 Extending the present design methodology to rank-deficient problems along the lines of \cite{JGLEE_almost} is left for future work.
 \end{remark}

\subsection{PageRank estimation} \label{subsection:PageRank}

PageRank is a widely used measure of nodal importance in directed networks, but each score depends on the global link structure. For a strongly connected digraph, consider the normalized PageRank score $r_i\ge0$ satisfying
\begin{equation}\label{eq:pagerank}
    r_i
    =
    \sum_{j\in\mathcal N_i}
    \frac{\alpha_{ij}}{d_j^{\text{out}}}r_j,
    \qquad
    \sum_{i=1}^{N}r_i=1,
\end{equation}
see, e.g., \cite{Page}. Let $r:=\operatorname{col}(r_i)_{i\in\mathcal{N}}$ and $D_{\text{out}}:=\operatorname{diag}(d_i^{\text{out}})_{i\in\mathcal{N}}$. Then \eqref{eq:pagerank} implies $\mathcal{L}_{\text{out}}D_{\text{out}}^{-1}r=0_N$.

Define $\sigma:=1_N^\top D_{\text{out}}^{-1}r$ and $p :=
    \frac{D_{\text{out}}^{-1}r}{\sigma}$.
Then $\mathcal{L}_{\text{out}}p=0_N$, $1_N^\top p=1$, and $p_i \sigma={r_i}/{d_i^{\text{out}}}$, so $p$ is the scaling vector in Section~\ref{subsection:special_3}. Now, consider the blended dynamics
\begin{equation}\label{eq:pagerank_blended}
    \dot{\mathsf{s}}(t)
    =
    -\frac{1}{\sigma}\mathsf{s}(t)+1,
\end{equation}
whose unique equilibrium is $\sigma$. Hence, to recover $r_i$, agent~$i$ must track the scaled value $p_i\sigma = r_i/ d_i^{\text{out}}$.

Under the design specialized in Section~\ref{subsection:special_3}, choose $f_1(x_1)=-d_1^{\text{out}}x_1+1$ and $f_i(x_i)=-d_i^{\text{out}}x_i$ for $i=2,\ldots,N$. Since $\sum_{i=1}^{N}p_i d_i^{\text{out}}=1/\sigma$, these choices give
$\sum_{i=1}^{N}f_i(p_i\mathsf{s})= - \sum_{i=1}^{N} p_i d_i^{\text{out}} \mathsf{s} +1= -\frac{1}{\sigma} \mathsf{s} + 1$, which gives \eqref{eq:pagerank_blended}. 

To obtain a direct estimate of $r_i$, one can simply redefine the state as $\bar x_i[n]:=d_i^{\text{out}}x_i[n]$ and let $\bar\delta_i:=d_i^{\text{out}}\delta_i$. Then, the implicit form of \eqref{eq:outdegree_algorithm} can be written explicitly as
\begin{equation}
\begin{aligned}
\bar x_i[n+1]
=
\frac{k}{k+{\bar\delta_i}}
\bigg( &
\frac{\bar\delta_i}{k} \mathbf{e}_1^\top \mathbf{e}_i 
+(1-\bar\delta_i)\bar x_i[n] \\
&~ +\bar\delta_i
\sum_{j\in\mathcal N_i}
\frac{\alpha_{ij}}{d_j^{\text{out}}}
\bar x_j[n]
\bigg).
\label{eq:pagerank_algorithm}
\end{aligned}
\end{equation}
Choose $0<\bar\delta_i\leq1$ for all $i\in\mathcal{N}$, with strict inequality for at least one agent. The algorithm is implemented locally by each agent~$j$ transmitting $\bar x_j[n]/d_j^{\text{out}}$ to its out-neighbors.

By Theorem~\ref{thm:uniform}, for sufficiently large $k$, $x_i[n]$ eventually remains close to $p_i\sigma=r_i/d_i^{\text{out}}$, and hence $\bar x_i[n]$ approximates $r_i$. Unlike the multi-step PageRank estimator in~\cite{JWKim}, \eqref{eq:pagerank_algorithm} applies the node and coupling updates at every iteration, without alternating between separate phases. It therefore avoids the need to align those phases across agents. Section~\ref{subsection:page} illustrates this difference through a comparison with misaligned node-update schedules in the multi-step method.

\section{Simulations} \label{sec:simulations}

\begin{figure}[!t]
\centering
\resizebox{0.5\linewidth}{!}{
\begin{tikzpicture}[
  >=Stealth,
  node/.style={
    circle, draw=black, line width=1.55pt,
    minimum size=7.3mm, inner sep=0pt,
    fill=gray!18,
    font=\fontsize{13.5}{13.5}\selectfont
  },
  anchornode/.style={node},
  edge/.style={
    -{Stealth[length=2.05mm,width=1.55mm]},
    line width=1.25pt,
    draw=black,
    shorten <=0.3pt,
    shorten >=0.3pt
  },
weight/.style={
  inner sep=0pt,
  font=\fontsize{10.5}{10.5}\selectfont
}
]
\begin{scope}
  \node[anchornode] (b1) at (0.0,  0.9) {1};
  \node[node]       (b2) at (2.4,  0.9) {2};
  \node[node]       (b3) at (4.8,  0.9) {3};
  \node[node]       (b4) at (3.6, -0.7) {4};
  \node[node]       (b5) at (1.2, -0.7) {5};

\draw[edge] (b1) -- (b2);
\draw[edge] (b2) -- (b3);
\draw[edge] (b2) -- (b4);
\draw[edge] (b3) -- (b4);
\draw[edge] (b4) -- (b5);
\draw[edge] (b2) -- (b5);
\draw[edge] (b5) -- (b1);

\node[weight] at (1.2,  1.2) {0.5};

\node[weight] at (3.6,  1.2) {0.6};

\node[weight] at (3.38, 0.10) {0.2};
\node[weight] at (4.58, 0.10) {1.1};

\node[weight] at (2.4, -1.0) {1};

\node[weight] at (1.42, 0.10) {0.5};
\node[weight] at (0.22, 0.10) {1.2};
\end{scope}
\end{tikzpicture}
}
\caption{Network topology for simulations.}
\label{fig1}
\end{figure}
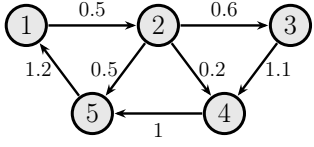

In this section, we illustrate the proposed framework through simulations of the $1/\zeta_i$ estimator, the distributed WLS solver, and the distributed PageRank algorithm in Section~\ref{section:applications}. Throughout this section, we use the strongly connected digraph with $N=5$ agents shown in Fig.~\ref{fig1}, where the edge weights are given in the figure.

\subsection{Online $1/\zeta_i$ estimation and weighted
least-squares}

Consider the WLS problem \eqref{eq:wls_problem} with
\begingroup
\renewcommand{\arraystretch}{0.85}
\[
\mathbf{A}=
\begin{bmatrix}
1 & 2 & 1 & 1\\
2 & -1 & -1 & 1\\
1 & -2 & 4 & -1\\
-1 & -0.6 & 0.4 & 1.8\\
1 & 2 & -5 & 1
\end{bmatrix},
\qquad
\mathbf{b}=
\begin{bmatrix}
1\\
2\\
1.5\\
1.7\\
1.1
\end{bmatrix}.
\]
\endgroup
We set $w_i=1$ for all $i\in\mathcal{N}$, for which the unique solution is
$x^*\approx[0.6491,\,-0.3223,\,0.1144,\,0.9795]^\top$. Each agent runs \eqref{eq:wls_distributed} with $\delta_i=1/(d_i+1)$. We also simulate the explicit counterpart obtained by evaluating the local gradient term at $x_i[n]$ instead of $x_i[n+1]$. 

For the online implementation, the estimator in Section~\ref{subsection:zeta} and the WLS algorithm are implemented simultaneously. At each iteration, agent~$i$ uses its current estimate of $1/\zeta_i$ in the WLS update. We use $k_\zeta$ for the parameter of the estimator and $k$ for that of the WLS algorithm.

\begin{figure}[!t]
    \centering
    \subfloat[Estimates of $1/\zeta_i$ ($k_\zeta=1$).]{%
        \includegraphics[width=0.49\columnwidth]{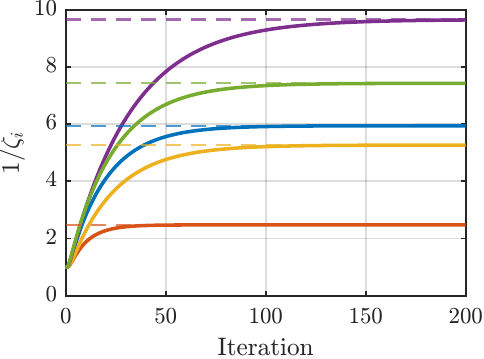}%
        \label{fig:zeta_estimation}%
    }%
    \hfill
    \subfloat[WLS solution error ($k=100$).]{%
        \includegraphics[width=0.49\columnwidth]{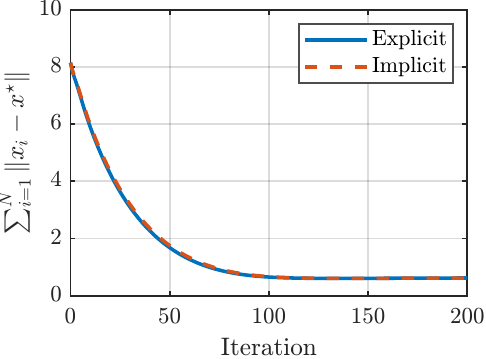}%
        \label{fig:wls_solution_error}%
    }%
    \caption{Online $1/\zeta_i$ estimation and WLS convergence. Dashed lines in (a) indicate the true values.}
    \label{fig:wls_results}
\end{figure}

\begin{figure}[!t]
        \centering
        \includegraphics[width=1\linewidth]{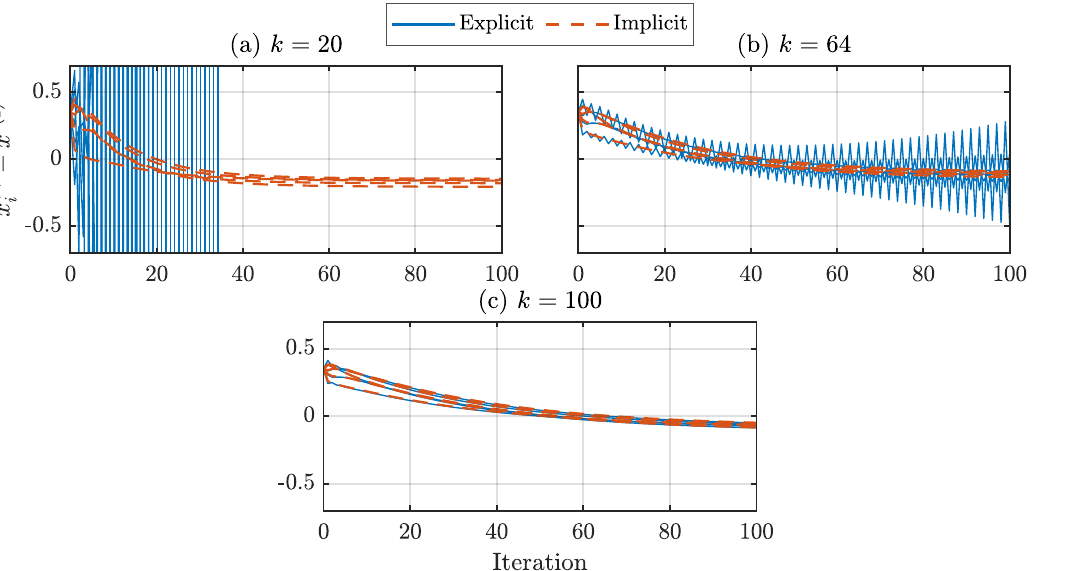}
        \caption{First components of $x_i[n]-x^*$ under the explicit and implicit WLS updates for
        (a) $k=20$, (b) $k=64$, and (c) $k=100$.}
        \label{fig:wls_explicit_implicit}
    \end{figure}

Fig.~\ref{fig:wls_results} shows the estimates of $1/\zeta_i$ for $k_\zeta=1$, and the WLS solution error $\sum_{i=1}^{N}\|x_i[n]-x^*\|$ for $k=100$. The estimates converge while the WLS error decreases, illustrating the concurrent online implementation in this example. As shown in Fig.~\ref{fig:wls_explicit_implicit}, the implicit update remains convergent for all three values, whereas the explicit update diverges for $k=20$, exhibits growing oscillations for $k=64$, and approaches the implicit behavior as $k$ increases.

    \subsection{PageRank algorithm in an asynchronous setting} \label{subsection:page}  
    
    We compare the proposed PageRank algorithm \eqref{eq:pagerank_algorithm} with the multi-step method in~\cite{JWKim}. For Fig.~\ref{fig1}, the PageRank scores are $r\approx[0.2453,\,0.2453,\,0.1132,\,0.1509,\,0.2453]^\top$. The multi-step method performs one node update every $K=50$ iterations and applies coupling updates otherwise. For the proposed algorithm, we set $k=50$ and ${\bar\delta}_i=0.8$ for all $i\in\mathcal{N}$. To represent an asynchronous setting for the multi-step method, the node-update times are shifted across agents so that the node updates are not performed simultaneously. The proposed algorithm has no separate node and coupling phases and applies the same update at every iteration.

\begin{figure}[!t]
    \centering
     {\includegraphics[width=0.9\linewidth]{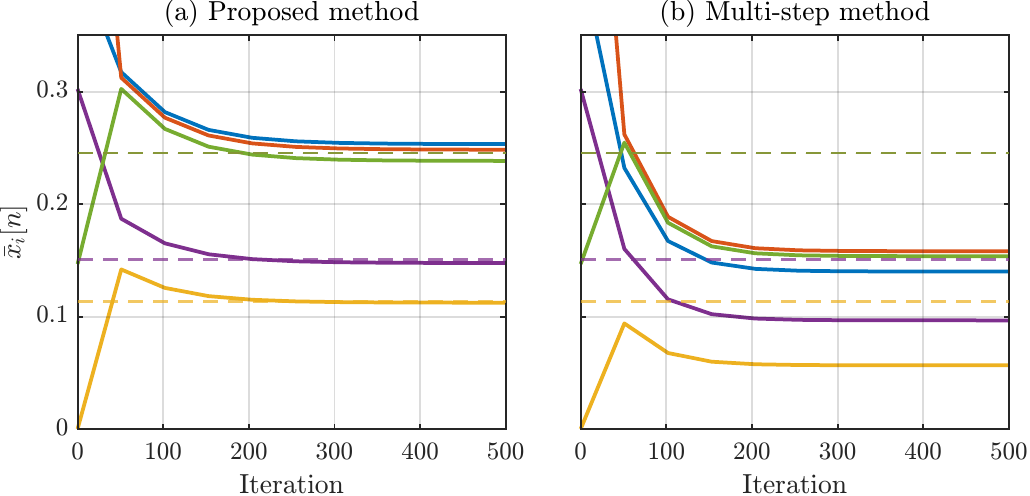}}
     
       \caption{PageRank estimation with asynchronous node updates. Solid lines represent $\bar x_i[n]$, and dashed lines indicate the PageRank scores.}
       \label{fig_pagerank_async}
    \end{figure}

    Fig.~\ref{fig_pagerank_async} shows the trajectories of $\bar x_i[n]$, with dashed lines indicating the PageRank scores. The proposed estimates eventually remain close to the PageRank scores, whereas those of the multi-step method do not when the node-update schedules are misaligned.

    \section{Conclusion} \label{sec:conclusion}

    This paper developed a decentralized design framework for discrete-time distributed algorithms based on their emergent blended dynamics. The analysis characterized the collective behavior on a rescaled time axis, established approximation bounds, and showed how prescribed blended dynamics can be realized through decentralized parameter choices. The framework was illustrated through four applications and also revealed the trade-off between approximation accuracy and iteration count. These results establish a general viewpoint for analyzing and designing discrete-time distributed algorithms through their emergent dynamics. Future work will extend this viewpoint to nonautonomous settings, broader stability conditions and coupling structures, and decentralized or adaptive tuning of $k$.

    \section*{Generative AI disclosure}
    ChatGPT assisted with language editing and manuscript organization. The authors reviewed the content and remain fully responsible for the manuscript.

\bibliographystyle{unsrt}
\bibliography{autosam}

@article{JKim,
  author = {J. Kim and J. Yang and H. Shim and J.-S. Kim and J. H. Seo},
  title = {Robustness of synchronization of heterogeneous agents by strong coupling and a large number of agents},
  journal = {IEEE Transactions on Automatic Control},
  volume = {61}, number = {10}, pages = {3096--3102}, year = {2016}, month = oct
}

@article{JGLEE_tool,
  author = {J. G. Lee and H. Shim},
  title = {A tool for analysis and synthesis of heterogeneous multi-agent systems under rank-deficient coupling},
  journal = {Automatica}, volume = {117}, pages = {108952}, year = {2020}
}

@article{JWKim,
  author = {J. W. Kim and J. G. Lee and D. Lee and H. Shim},
  title = {A design method of distributed algorithms via discrete-time blended dynamics theorem},
  journal = {Automatica}, volume = {159}, pages = {111371}, year = {2024}, month = jan
}

@inproceedings{LWang,
  author = {L. Wang and J. Liu and A. S. Morse and B. D. O. Anderson},
  title = {A distributed observer for a discrete-time linear system},
  booktitle = {Proceedings of the IEEE Conference on Decision and Control (CDC)},
  pages = {6381--6386}, year = {2019}
}

@article{Khalil,
  author = {B. Litkouhi and H. K. Khalil},
  title = {Infinite-time regulators for singularly perturbed difference equations},
  journal = {International Journal of Control},
  volume = {39}, number = {3}, pages = {587--598}, year = {1984}
}

@article{DZhang,
  author = {D. Zhang and L. Liu and G. Feng},
  title = {Consensus of heterogeneous linear multiagent systems subject to aperiodic sampled-data and {DoS} attack},
  journal = {IEEE Transactions on Cybernetics},
  volume = {49}, number = {4}, pages = {1501--1511}, year = {2019}, month = apr
}

@article{SLi,
  author = {S. Li and G. Feng and X. Luo and X. Guan},
  title = {Output consensus of heterogeneous linear discrete-time multiagent systems with structural uncertainties},
  journal = {IEEE Transactions on Cybernetics},
  volume = {45}, number = {12}, pages = {2868--2879}, year = {2015}, month = dec
}

@article{XXu,
  author = {X. Xu and L. Liu and G. Feng},
  title = {Consensus of discrete-time linear multiagent systems with communication, input and output delays},
  journal = {IEEE Transactions on Automatic Control},
  volume = {63}, number = {2}, pages = {492--497}, year = {2018}, month = feb
}

@article{YYan,
  author = {Y. Yan and Z. Chen and R. H. Middleton},
  title = {Autonomous synchronization of heterogeneous multiagent systems},
  journal = {IEEE Transactions on Control of Network Systems},
  volume = {8}, number = {2}, pages = {940--950}, year = {2021}, month = jun
}

@incollection{JGLEE_design,
  author = {J. G. Lee and H. Shim},
  title = {Design of heterogeneous multi-agent systems for distributed computation},
  booktitle = {Trends in Nonlinear and Adaptive Control: A Tribute to Laurent Praly for His 65th Birthday},
  pages = {83--108}, year = {2021}
}

@article{SLee,
  author = {S. Lee and H. Shim},
  title = {Blended dynamics approach to distributed optimization: Sum convexity and convergence rate},
  journal = {Automatica}, volume = {141}, pages = {110290}, year = {2022}
}

@article{TKim,
  author = {T. Kim and C. Lee and H. Shim},
  title = {Completely decentralized design of distributed observer for linear systems},
  journal = {IEEE Transactions on Automatic Control},
  volume = {65}, number = {11}, pages = {4664--4678}, year = {2020}, month = nov
}

@article{HYun,
  author = {H. Yun and H. Shim and H.-S. Ahn},
  title = {Initialization-free privacy-guaranteed distributed algorithm for economic dispatch problem},
  journal = {Automatica}, volume = {102}, pages = {86--93}, year = {2019}, month = apr
}

@book{Bullo,
  author = {F. Bullo},
  title = {Lectures on Network Systems},
  publisher = {Kindle Direct Publishing},
  volume = {1}, address = {Seattle, WA, USA}, 
  year = {2020}
}

@article{Slotine,
  author = {W. Lohmiller and J.-J. E. Slotine},
  title = {On contraction analysis for nonlinear systems},
  journal = {Automatica}, volume = {34}, number = {6}, pages = {683--696}, year = {1998}
}

@inproceedings{DGLee,
  author = {D. Lee and S. Lee and T. Kim and H. Shim},
  title = {Distributed algorithm for the network size estimation: Blended dynamics approach},
  booktitle = {Proceedings of the IEEE Conference on Decision and Control (CDC)},
  year = {2018}
}

@article{APriolo,
  author = {A. Priolo and A. Gasparri and E. Montijano and E. Sagu{\'e}s},
  title = {A distributed algorithm for average consensus on strongly connected weighted digraphs},
  journal = {Automatica}, volume = {50}, number = {3}, pages = {939--946}, year = {2014}
}

@article{JGLEE_almost,
  author = {J. G. Lee and H. Shim},
  title = {A distributed algorithm that finds almost best possible estimate under non-vanishing and time-varying measurement noise},
  journal = {IEEE Control Systems Letters}, volume = {4}, number = {1}, pages = {229--234}, year = {2020}, month = jan
}

@article{Page,
  author = {S. Brin and L. Page},
  title = {The anatomy of a large-scale hypertextual {Web} search engine},
  journal = {Computer Networks and ISDN Systems},
  volume = {30}, number = {1--7}, pages = {107--117}, year = {1998}
}

\appendix
\setcounter{lemma}{0}
\renewcommand{\thelemma}{\Alph{section}.\arabic{lemma}}

\section{Proofs of Theorems~\ref{thm:lim_sol}--\ref{thm:iteration}}

For brevity, we write $x[n]$ and $s[n]$ for $x[n,k]$ and $s[n,k]$, respectively, when no ambiguity arises. The notation $[\,\cdot\,]$ denotes $n$ for the explicit update and $n+1$ for the implicit update. 

\subsection{Synchronization error estimates}
\label{app:sync-comparison}

We first derive synchronization error estimates used in the proofs of Theorems~\ref{thm:lim_sol} and~\ref{thm:uniform}, using the notation of Theorem~\ref{thm:uniform}. The coordinate decomposition gives
\begin{equation}
x_i[n]-p_i s[n] = p_i\left(\xi_1[n]-s[n]\right) + (R_i\otimes I_m)\tilde\xi[n].
\label{eq:app-error-decomposition}
\end{equation}

Let $\mathsf H=I_m$ for Theorem~\ref{thm:lim_sol} and $\mathsf H=H$ for Theorem~\ref{thm:uniform}, and define
\[
V[n]:=\|\mathsf H(\xi_1[n]-s[n])\|,
\quad
W[n]:=\|(\mathsf T R)_{\otimes m}\tilde\xi[n]\|.
\]
Then \eqref{eq:app-error-decomposition} yields
\begin{equation}
\|x_i[n]-p_i s[n]\|
\le
p_i\|\mathsf H^{-1}\|V[n]
+
\|\mathsf T^{-1}\|W[n].
\label{eq:app-reconstruction}
\end{equation}

Since $f_{\mathsf{s}}(z)=(q^\top A)_{\otimes m}F(p_{\otimes m}z)$, subtracting \eqref{eq:emer_dyn} from \eqref{eq:xi_1_dynamics} and taking norms gives, for either update,
\begin{equation}\label{eq:app-slow-error}
\begin{aligned}
&V[n+1] =  \left\|\mathsf H (\xi_1[n+1]-s[n+1]) \right\| \\
& \le \Bigl\|\mathsf H \Bigl\{ \xi_1[n]-s[n] + \frac{1}{k} \bigl( f_{\mathsf{s}}(\xi_1[\,\cdot\,])-f_{\mathsf{s}}(s[\,\cdot\,]) \bigr)\Bigr\} \Bigr\| \\
& \quad + \frac{\|\mathsf H\|\|q^\top A\|}{k}
\bigl\|F(x[\,\cdot\,])
-F(p_{\otimes m}\xi_1[\,\cdot\,])\bigr\|.
\end{aligned}
\end{equation}
Similarly, using $R\Lambda=(\mathcal{W}-\mathsf P)R$ and $RZ^\top=I_N-\mathsf P$ in \eqref{eq:xi_tilde_dynamics} gives
\begin{equation}\label{eq:app-disagreement-error}
\begin{aligned}
 W[n+1]   = {} & \|(\mathsf T R)_{\otimes m}\tilde\xi[n+1]\| \\
\le {} & \mu_{\mathcal{W}} W[n] +\frac{d_A}{k} \left\|F(x[\,\cdot\,])-F(p_{\otimes m}s[\,\cdot\,]) \right\| \\
& +\frac{d_A}{k}\left\|F(p_{\otimes m}s[\,\cdot\,]) \right\|,
\end{aligned}
\end{equation}
where $d_A:=\|\mathsf T(I_N-\mathsf P)A\|$.

These bounds can be simplified using
\begin{equation}\label{eq:app-F-bounds}
\begin{aligned}
&\|F(x[\,\cdot\,])-F(p_{\otimes m}\xi_1[\,\cdot\,])\|
\le L\|\mathsf T^{-1}\|W[\,\cdot\,],\\
&\|F(x[\,\cdot\,])-F(p_{\otimes m}s[\,\cdot\,])\|\\
&\qquad\le
L\|p\|\|\mathsf H^{-1}\|V[\,\cdot\,]
+L\|\mathsf T^{-1}\|W[\,\cdot\,],\\
&\|F(p_{\otimes m}s[\,\cdot\,])\|\le
L\|p\|\|\mathsf H^{-1}\|\|\mathsf Hs[\,\cdot\,]\|
+\sqrt N M_f,
\end{aligned}
\end{equation}
where $M_f:=\max_{i\in\mathcal{N}}\|f_i(0)\|$. 

We next estimate the first norm on the right-hand side of
\eqref{eq:app-slow-error}.
For $z_1,z_2\in\mathbb{R}^m$, let $z=z_1-z_2$ and
$h=f_{\mathsf{s}}(z_1)-f_{\mathsf{s}}(z_2)$.
Lipschitz continuity gives
$\|z+h/k\|\le(1+L_{\mathsf{s}}/k)\|z\|$ and
$\|z-h/k\|\ge(1-L_{\mathsf{s}}/k)\|z\|$.

Under Assumption~\ref{assum:contractive}, define
\begin{subequations}\label{eq:app-eta}
\begin{align}
\eta_k^{\rm E}
&:=\sqrt{1-\frac{2\rho_{\mathsf{s}}}{k}
+\frac{(L_{\mathsf{s}}\|H\|\|H^{-1}\|)^2}{k^2}},
\label{eta_exp}\\
\eta_k^{\rm I}
&:=\frac{1}{1+\rho_{\mathsf{s}}/k}.
\label{eta_imp}
\end{align}
\end{subequations}
Indeed, expanding the square gives 
$\|H(z+h/k)\|^2\le(\eta_k^{\rm E})^2\|Hz\|^2$, while contractivity gives
$z^\top H^2(z-h/k)\ge(1+\rho_{\mathsf{s}}/k)\|Hz\|^2$.
Taking square roots in the first inequality and using
Cauchy--Schwarz in the second yields
\begin{subequations}\label{eq:app-map-estimates}
\begin{align}
\|H(z+h/k)\|
&\le\eta_k^{\rm E}\|Hz\|,
\label{eq:app-map-E}\\
\|H(z-h/k)\|
&\ge(\eta_k^{\rm I})^{-1}\|Hz\|.
\label{eq:app-map-I}
\end{align}
\end{subequations}
Here, $\eta_k^{\rm E}$ is real since
$0<\rho_{\mathsf{s}}\le L_{\mathsf{s}}\|H\|\|H^{-1}\|$;
moreover, $\eta_k^{\rm E}\in[0,1)$ for $k>\bar k$ and
$\eta_k^{\rm I}\in(0,1)$ for $k>0$.

In the comparisons below, $a_k^{\rm E}$ and $a_k^{\rm I}$
denote the coefficients used to bound $V[n+1]$.
For Theorem~\ref{thm:lim_sol}, take
$a_k^{\rm E}=1+L_{\mathsf{s}}/k$ and
$a_k^{\rm I}=(1-L_{\mathsf{s}}/k)^{-1}$,
with $k>L_{\mathsf{s}}$ in the implicit case;
for Theorem~\ref{thm:uniform}, take
$a_k^{\rm E}=\eta_k^{\rm E}$ and
$a_k^{\rm I}=\eta_k^{\rm I}$.
Applying the corresponding estimates to
\eqref{eq:app-slow-error}--\eqref{eq:app-F-bounds}
gives the following comparisons.

    \begin{subequations}\label{eq:app-sync-comparisons}
    For the explicit update,
    \begin{equation}\label{eq:app-sync-comparison-E}
    \begin{aligned}
    \begin{bmatrix}V[n+1]\\W[n+1]\end{bmatrix}
    \le{}&\mathcal M_k^{\rm E}
           \begin{bmatrix}V[n]\\W[n]\end{bmatrix}+\frac{d_A}{k}\|F(p_{\otimes m}s[n])\|
           \begin{bmatrix}0\\1\end{bmatrix},
    \end{aligned}
    \end{equation}
    where
    \[
    \mathcal M_k^{\rm E}
    :=
    \begin{bmatrix}
    a_k^{\rm E} & \ell_1/k\\
    \ell_2/k & \mu_{\mathcal{W}}+\ell_3/k
    \end{bmatrix},
    \]
    with $\ell_1:=L\|\mathsf H\|\|q^\top A\|\|\mathsf T^{-1}\|$, $\ell_2:=Ld_A\|p\|\|\mathsf H^{-1}\|$, and $\ell_3:=Ld_A\|\mathsf T^{-1}\|$.
        
    	For the implicit update, $y\mapsto\mathcal{W}_{\otimes m}x[n]+A_{\otimes m}F(y)/k$ is a contraction when $k>L\|A\|$, so the update is uniquely solvable. Take $z=\xi_1[n+1]-s[n+1]$ and $h=f_{\mathsf{s}}(\xi_1[n+1])-f_{\mathsf{s}}(s[n+1])$. Subtracting \eqref{eq:emer_dyn_implicit} from \eqref{eq:xi_1_dynamics} and moving $h/k$ to the left before taking norms gives
        \begin{align*}
			(a_k^{\rm I})^{-1}V[n+1] \le\|\mathsf H(z-h/k)\|\le V[n]+\frac{\ell_1}{k}W[n+1].
		\end{align*}
		Also, \eqref{eq:app-disagreement-error} and	\eqref{eq:app-F-bounds} give
		\begin{align*}
			\left(1-\frac{\ell_3}{k}\right)W[n+1]
			&\le\mu_{\mathcal{W}} W[n]+\frac{\ell_2}{k}V[n+1]\\
			&\quad+\frac{d_A}{k}\|F(p_{\otimes m}s[n+1])\|.
		\end{align*}
		For all sufficiently large $k$, $(a_k^{\rm I})^{-1}>0$,	$1-\ell_3/k>0$, and
		$\omega_k:=(a_k^{\rm I})^{-1}\left(1-{\ell_3}/{k}\right) -{\ell_1\ell_2}/{k^2}>0$.
		The coefficient matrix of $V[n+1],W[n+1]$ then has a nonnegative inverse. Multiplying the two inequalities by this inverse gives
		\begin{equation}\label{eq:app-sync-comparison-I}
			\begin{aligned}
				\begin{bmatrix}V[n+1]\\W[n+1]\end{bmatrix}
				\le{}&\mathcal M_k^{\rm I}
				\begin{bmatrix}V[n]\\W[n]\end{bmatrix}\\
				&+\frac{d_A\|F(p_{\otimes m}s[n+1])\|}{k\omega_k}
				\begin{bmatrix}\ell_1/k\\ (a_k^{\rm I})^{-1}\end{bmatrix},
			\end{aligned}
		\end{equation}
		where
		\[
		\mathcal M_k^{\rm I}:=
		\frac1{\omega_k}
		\begin{bmatrix}
			1-\ell_3/k & \mu_{\mathcal{W}}\ell_1/k\\
			\ell_2/k & \mu_{\mathcal{W}}/a_k^{\rm I}
		\end{bmatrix}.
		\]
\end{subequations}

\subsection{Proof of Theorem~\ref{thm:lim_sol}}
\label{app:proof-theorem1}

Let $s[\cdot]$ solve \eqref{eq:emer_dyn} with $s[0]=\mathsf{s}(0)=q^\top_{\otimes m}x[0]$. By Lemma~\ref{lem:discret_error}, for each $T>0$, it is enough to show
\begin{equation}
 \lim_{k\to\infty}
 \|x_i[n_{T,k},k]-p_i s[n_{T,k},k]\|=0, \qquad i\in\mathcal{N}.
\label{eq:app-T1-goal}
\end{equation}
We first use the following finite-horizon bound.
\begin{lemma}
\label{lem:app-finite-horizon-s}
Let $s[n,k]$ be the solution of \eqref{eq:emer_dyn} with initial condition $s[0]$, under either update. For every $T>0$, there exists $S_T>0$ such that $\|s[n,k]\|\le S_T$ for $0\le n\le n_{T,k}$ and all sufficiently large $k$.
\end{lemma}
\begin{proof}
For $k>L_{\mathsf{s}}$, Lipschitz continuity and
$1+L_{\mathsf{s}}/k\le(1-L_{\mathsf{s}}/k)^{-1}$ give, for either update,
\[
\|s[n+1]\|
\le
\frac{\|s[n]\|+\|f_{\mathsf{s}}(0)\|/k}{1-L_{\mathsf{s}}/k}.
\]
Iteration yields
\[
\|s[n]\|
\le
(1-L_{\mathsf{s}}/k)^{-n_{T,k}}
\left(\|s[0]\|+\frac{n_{T,k}}{k}\|f_{\mathsf{s}}(0)\|\right)
\]
for $0\le n\le n_{T,k}$.
Since $n_{T,k}/k\to T/\tau_{\mathsf{s}}$ and
$(1-L_{\mathsf{s}}/k)^{-n_{T,k}}\to e^{L_{\mathsf{s}}T/\tau_{\mathsf{s}}}$,
the right-hand side is uniformly bounded for large $k$.
\end{proof}

For Theorem~\ref{thm:lim_sol}, take $\mathsf H=I_m$. Since $s[0]=\xi_1[0]$, we have $V[0]=0$. By Lemma~\ref{lem:app-finite-horizon-s} and \eqref{eq:app-F-bounds}, there exists $B_T>0$ such that $d_A\|F(p_{\otimes m}s[n])\|\le B_T$ for $0\le n\le n_{T,k}$ and all sufficiently large $k$.

In the implicit case, $\omega_k=(1-L_{\mathsf{s}}/k)(1-\ell_3/k)-\ell_1\ell_2/k^2\to1$ and $(1-\ell_3/k)/\omega_k=1+[L_{\mathsf{s}}(1-\ell_3/k)+\ell_1\ell_2/k]/(k\omega_k)$. Thus, the first diagonal entry of $\mathcal M_k^{\rm I}$ is $1+O(k^{-1})$, its off-diagonal	entries are $O(k^{-1})$, and its second diagonal entry tends to $\mu_{\mathcal{W}}$. By the bound $B_T$, the two components of the additional vector in \eqref{eq:app-sync-comparison-I} are	$O(k^{-2})$ and $O(k^{-1})$, respectively, uniformly for $0\le n<n_{T,k}$.

Choose $\bar\mu\in(\mu_{\mathcal{W}},1)$. Using these estimates in \eqref{eq:app-sync-comparison-E} and \eqref{eq:app-sync-comparison-I} gives $C_T>0$, independent of $n,k$, such that, for all sufficiently large $k$ and $0\le n<n_{T,k}$,
\begin{align}
V[n+1]
&\le
\left(1+\frac{C_T}{k}\right)V[n]
+\frac{C_T}{k}W[n]
+\frac{C_T}{k^2},
\label{eq:app-T1-rec-V}\\
W[n+1]
&\le
\bar\mu W[n]
+\frac{C_T}{k}V[n]
+\frac{C_T}{k}.
\label{eq:app-T1-rec-W}
\end{align}
The $k^{-2}$ term appears only in the implicit case.

Let $V_T:=\max_{0\le j\le n_{T,k}}V[j]$. Iterating \eqref{eq:app-T1-rec-W} gives, for $0\le n\le n_{T,k}$,
\begin{align} \label{eq:app-T1-W-bound}
W[n]
&\le\bar\mu^nW[0] +\frac{C_T}{k}\sum_{j=0}^{n-1} \bar\mu^{n-1-j}(1+V[j]) \notag \\
&\le\bar\mu^nW[0] + \frac{C_T}{k(1-\bar\mu)}(1+V_T).
\end{align}
Since $V[0]=0$, substituting \eqref{eq:app-T1-W-bound} into \eqref{eq:app-T1-rec-V} and iterating gives
\begin{align*}
V[n]\le{}&
\frac{C_TW[0]}{k}
\sum_{j=0}^{n-1} \bar\mu^j \left(1+\frac{C_T}{k}\right)^{n-1-j}\\
&+
\left( \frac{C_T^2(1+V_T)}{k^2(1-\bar\mu)}
+\frac{C_T}{k^2} \right)
\sum_{j=0}^{n-1}
\left(1+\frac{C_T}{k}\right)^{n-1-j}.
\end{align*}
The two sums are bounded by $e^{C_Tn_{T,k}/k}/(1-\bar\mu)$ and $n_{T,k}e^{C_Tn_{T,k}/k}$, respectively. Since $n_{T,k}/k$ remains bounded as $k\to\infty$, taking the maximum over $0\le n\le n_{T,k}$ gives $V_T\le C_T'(1+W[0]+V_T)/k$ for some $C_T'>0$ independent of $k$. Thus, for $k\ge2C_T'$, $V_T\le2C_T'(1+W[0])/k$, and hence $V_T\to0$. Since $n_{T,k}\to\infty$ and $\bar\mu<1$, \eqref{eq:app-T1-W-bound} then gives $W[n_{T,k}]\to0$. Equation~\eqref{eq:app-reconstruction} therefore proves \eqref{eq:app-T1-goal}, and Lemma~\ref{lem:discret_error} completes the proof.

\subsection{Proof of Theorem~\ref{thm:uniform}}
\label{app:proof-theorem2}

We first split the error as
\begin{equation}
\begin{aligned}
&\|x_i[n,k]-p_i\mathsf{s}(t_{n,k})\|\\
&\quad\le
\|x_i[n,k]-p_i s[n,k]\|
+p_i\|s[n,k]-\mathsf{s}(t_{n,k})\|.
\end{aligned}
\label{eq:app-error-split}
\end{equation}

To bound the synchronization term, we first estimate $\|Hs[n,k]\|$.
\begin{lemma}
\label{lem:app-s-bound}
Under Assumption~\ref{assum:contractive}, let $s[n,k]$ be the solution of \eqref{eq:emer_dyn}. For $k>\bar k$ in the explicit case and $k>0$ in the implicit case, $\|Hs[n,k]\|
\le s_{\infty,k}^r+(\eta_k^r)^n\Delta_s^r$ for all $n\ge0$ and $r\in\{\mathrm E,\mathrm I\}$,
where $s_{\infty,k}^{\rm E}:=\|Hf_{\mathsf{s}}(0)\|/[\rho_{\mathsf{s}}(1-\bar k/k)]$,
$s_{\infty,k}^{\rm I}:=\|Hf_{\mathsf{s}}(0)\|/\rho_{\mathsf{s}}$, and
$\Delta_s^r:=(\|Hs[0]\|-s_{\infty,k}^r)_+$.
\end{lemma}
\begin{proof}
For each $k>0$, Assumption~\ref{assum:contractive} and Lipschitz continuity imply that $y\mapsto y-\alpha(y-s[n]-f_{\mathsf{s}}(y)/k)$ is a contraction in the norm $\|H\cdot\|$ for sufficiently small $\alpha>0$. Its unique fixed point solves \eqref{eq:emer_dyn_implicit}.

Applying \eqref{eq:app-map-estimates} with one argument equal to zero to \eqref{eq:emer_dyn} gives, respectively,
\begin{align*}
\|Hs[n+1]\|&\le\eta_k^{\rm E}\|Hs[n]\|+\|Hf_{\mathsf{s}}(0)\|/k,\\
\|Hs[n+1]\|&\le\eta_k^{\rm I}\|Hs[n]\|
                  +\eta_k^{\rm I}\|Hf_{\mathsf{s}}(0)\|/k.
\end{align*}
For $k>\bar k$, we have
\begin{equation}\label{eq:one-minus-eta-E-full}
\frac{\rho_{\mathsf{s}}(k-\bar k)}{k^2}
\le 1-\eta_k^{\rm E}
=\frac{2\rho_{\mathsf{s}}(k-\bar k)}{k^2(1+\eta_k^{\rm E})}
\le\frac{2\rho_{\mathsf{s}}}{k}.
\end{equation}
Together with $\eta_k^{\rm I}/(1-\eta_k^{\rm I})=k/\rho_{\mathsf{s}}$, this implies that the constant term in each inequality is at most $(1-\eta_k^r)s_{\infty,k}^r$. Hence $(\|Hs[n+1]\|-s_{\infty,k}^r)_+ \le\eta_k^r(\|Hs[n]\|-s_{\infty,k}^r)_+$. Iteration proves the result.
\end{proof}

For Theorem~\ref{thm:uniform}, take $\mathsf H=H$. By Lemma~\ref{lem:app-s-bound} and \eqref{eq:app-F-bounds},
\begin{subequations}
\label{eq:app-reduced-input}
\begin{align}
\frac{d_A}{k}\|F(p_{\otimes m}s[n])\|
&\le
\frac{\ell_2}{k}(\eta_k^{\rm E})^n\Delta_s^{\rm E}
+g_k^{\rm E},
\\
\frac{d_A}{k}\|F(p_{\otimes m}s[n+1])\|
&\le
\frac{\ell_2}{k}(\eta_k^{\rm I})^{n+1}\Delta_s^{\rm I}
+g_k^{\rm I},
\end{align}
\end{subequations}
where
$g_k^r:=d_A(L\|p\|\|H^{-1}\|s_{\infty,k}^r+\sqrt N M_f)/k$.

We next establish Schur stability of the matrices $\mathcal M_k^{\rm E}$ and $\mathcal M_k^{\rm I}$ in \eqref{eq:app-sync-comparison-E}--\eqref{eq:app-sync-comparison-I}. Since $\mathsf P=pq^\top$, $\|\mathsf P\|=\|p\|\|q\|\ge1$ and
$\|\mathsf T(I_N-\mathsf P)\|\|\mathsf T^{-1}\|\ge1$. Thus $\kappa_{\mathcal{W}}\ge1$ and $|\lambda_2|\le\mu_{\mathcal{W}}\le\lambda_{\mathcal{W}}<1$. Since $A$ is diagonal, the definitions give $\ell_3\le\kappa_{\mathcal{W}} L\|A\|$ and $\ell_1\ell_2\le\kappa_{\mathcal{W}} L^2\|A^2\|\|H\|\|H^{-1}\|$.
Since $1-\mu_{\mathcal{W}}=\kappa_{\mathcal{W}}(1-\lambda_{\mathcal{W}})$, \eqref{coupling_gain_lower_bound} implies
\begin{equation}
\rho_{\mathsf{s}}\ell_3+\ell_1\ell_2 \le\frac{\rho_{\mathsf{s}}(1-\mu_{\mathcal{W}})}2 k^*.
\label{eq:app-kstar-control}
\end{equation}
For $k>k^*$, $\ell_3/k<(1-\mu_{\mathcal{W}})/2$ and $\ell_1\ell_2/k^2<\rho_{\mathsf{s}}(1-\mu_{\mathcal{W}})/(2k)$. For the implicit update, these inequalities give
\[
\omega_k>
\left(1+\frac{\rho_{\mathsf{s}}}{k}\right)\frac{1+\mu_{\mathcal{W}}}{2} -\frac{\rho_{\mathsf{s}}(1-\mu_{\mathcal{W}})}{2k}>0.
\]
Direct expansion gives $\det(I_2-\mathcal M_k^{\rm E})=(1-\eta_k^{\rm E})(1-\mu_{\mathcal W}-\ell_3/k)-\ell_1\ell_2/k^2$ and $\omega_k\det(I_2-\mathcal M_k^{\rm I})=(\rho_{\mathsf s}/k)(1-\mu_{\mathcal W}-\ell_3/k)-\ell_1\ell_2/k^2$. Using \eqref{eq:one-minus-eta-E-full}, $(k-\bar k)/k\le1$, and \eqref{eq:app-kstar-control} yields
\begin{equation}\label{eq:app-determinants}
\begin{aligned}
\det(I_2-\mathcal M_k^{\rm E})
&\ge
\frac{\rho_{\mathsf{s}}(1-\mu_{\mathcal{W}})}{k^2}
(k-\bar k-k^*),
\\
\omega_k\det(I_2-\mathcal M_k^{\rm I})
&\ge
\frac{\rho_{\mathsf{s}}(1-\mu_{\mathcal{W}})}{k^2}
(k-k^*).
\end{aligned}
\end{equation}
The bounds are positive for $k>\bar k+k^*$ in the explicit case and $k>k^*$ in the implicit case. Both matrices are nonnegative with first diagonal entries less than one: $\eta_k^{\rm E}<1$, and
\[
\omega_k-(1-\ell_3/k)
=\omega_k\det(I_2-\mathcal M_k^{\rm I})+\rho_{\mathsf{s}}\mu_{\mathcal{W}}/k>0.
\]
For a nonnegative $M=\left[\begin{smallmatrix}a&b\\c&d\end{smallmatrix}\right]$ with $a<1$, the condition $\det(I_2-M)>0$ gives $(1-a)(1-d)>bc\ge0$, so $d<1$ and $\rho(M)=(a+d+\sqrt{(a-d)^2+4bc})/2<1$. Thus $\rho(\mathcal M_k^r)<1$.

Since $\mathcal M_k^r$ is Schur stable, we define $(V_{\infty,k}^r,W_{\infty,k}^r)$ from the $n$-independent terms in \eqref{eq:app-reduced-input} by
\begin{subequations}
\label{eq:app-steady-systems}
\begin{align}
(I_2-\mathcal M_k^{\rm E})
\begin{bmatrix}
V_{\infty,k}^{\rm E}\\
W_{\infty,k}^{\rm E}
\end{bmatrix}
&=
\begin{bmatrix}
0\\
g_k^{\rm E}
\end{bmatrix},
\\
(I_2-\mathcal M_k^{\rm I})
\begin{bmatrix}
V_{\infty,k}^{\rm I}\\
W_{\infty,k}^{\rm I}
\end{bmatrix}
&=
\frac{g_k^{\rm I}}{\omega_k}
\begin{bmatrix}
\ell_1/k\\
1+\rho_{\mathsf{s}}/k
\end{bmatrix}.
\end{align}
\end{subequations}
Solving \eqref{eq:app-steady-systems} gives
\begin{subequations}
\label{eq:app-steady-solutions}
\begin{align}
V_{\infty,k}^{\rm E}
&=
\frac{(\ell_1/k)g_k^{\rm E}}
{\det(I_2-\mathcal M_k^{\rm E})},
&
W_{\infty,k}^{\rm E}
&=
\frac{(1-\eta_k^{\rm E})g_k^{\rm E}}
{\det(I_2-\mathcal M_k^{\rm E})},
\\
V_{\infty,k}^{\rm I}
&=
\frac{(\ell_1/k)g_k^{\rm I}}
{\omega_k\det(I_2-\mathcal M_k^{\rm I})},
&
W_{\infty,k}^{\rm I}
&=
\frac{(\rho_{\mathsf{s}}/k)g_k^{\rm I}}
{\omega_k\det(I_2-\mathcal M_k^{\rm I})}.
\end{align}
\end{subequations}
Combining \eqref{eq:app-steady-solutions} with \eqref{eq:app-determinants}, there exist constants $\beta_2^{\rm E},\beta_2^{\rm I}>0$ and $\beta_3^{\rm E}\ge0$ such that, for every $i\in\mathcal{N}$,
\begin{subequations}
\label{eq:app-steady-bounds}
\begin{align}
\begin{split}
& p_i\|H^{-1}\|V_{\infty,k}^{\rm E}
+\|\mathsf T^{-1}\|W_{\infty,k}^{\rm E}\\  
& \qquad \le
\frac{\beta_2^{\rm E}}{k-\bar k-k^*} 
+\frac{\beta_3^{\rm E}}{(k-\bar k-k^*)^2},
\end{split}
\\
& p_i\|H^{-1}\|V_{\infty,k}^{\rm I}
+\|\mathsf T^{-1}\|W_{\infty,k}^{\rm I}
\le \frac{\beta_2^{\rm I}}{k-k^*}.
\end{align}
\end{subequations}
One choice is obtained as follows. Let $S_{\mathsf{s}}:=\|Hf_{\mathsf{s}}(0)\|/\rho_{\mathsf{s}}$, $B_0:=L\|p\|\|H^{-1}\|S_{\mathsf{s}}+\sqrt N M_f$, and $B_1:=L\|p\|\|H^{-1}\|S_{\mathsf{s}}\bar k$, and define
\begin{align*}
C^{\rm E}&:=
\frac{d_A\bigl(p_{\max}\|H^{-1}\|\ell_1
                 +2\rho_{\mathsf{s}}\|\mathsf T^{-1}\|\bigr)}
     {\rho_{\mathsf{s}}(1-\mu_{\mathcal{W}})},\\
C^{\rm I}&:=
\frac{d_A\bigl(p_{\max}\|H^{-1}\|\ell_1
                 +\rho_{\mathsf{s}}\|\mathsf T^{-1}\|\bigr)}
     {\rho_{\mathsf{s}}(1-\mu_{\mathcal{W}})}.
\end{align*}
For $z=(k-\bar k-k^*)^{-1}$, $s_{\infty,k}^{\rm E}=S_{\mathsf{s}}(1+\bar k/(k-\bar k)) \le S_{\mathsf{s}}(1+\bar k z)$, so $g_k^{\rm E}\le d_A(B_0+B_1z)/k$; also $g_k^{\rm I}=d_AB_0/k$. Equations~\eqref{eq:one-minus-eta-E-full}, \eqref{eq:app-steady-solutions}, and \eqref{eq:app-determinants} therefore bound the left-hand sides of \eqref{eq:app-steady-bounds} by $C^{\rm E}(B_0z+B_1z^2)$ and $C^{\rm I}B_0/(k-k^*)$, respectively. Thus we may take $\beta_2^{\rm E}:=1+C^{\rm E}B_0$, $\beta_3^{\rm E}:=C^{\rm E}B_1$, and $\beta_2^{\rm I}:=1+C^{\rm I}B_0$, which satisfy $\beta_2^{\rm E},\beta_2^{\rm I}>0$ and $\beta_3^{\rm E}\ge0$.

To bound the deviations, define $\hat V^r[n]:=(V[n]-V_{\infty,k}^r)_+$ and $\hat W^r[n]:=(W[n]-W_{\infty,k}^r)_+$. Subtracting \eqref{eq:app-steady-systems} from \eqref{eq:app-sync-comparison-E}--\eqref{eq:app-sync-comparison-I} and using \eqref{eq:app-reduced-input} and the nonnegativity of $\mathcal M_k^r$ gives
\begin{subequations}
\label{eq:app-transient-comparison}
\begin{align}
\begin{bmatrix}
\hat V^{\rm E}[n+1]\\
\hat W^{\rm E}[n+1]
\end{bmatrix}
&\le
\mathcal M_k^{\rm E}
\begin{bmatrix}
\hat V^{\rm E}[n]\\
\hat W^{\rm E}[n]
\end{bmatrix}
+
\frac{\ell_2}{k}
(\eta_k^{\rm E})^n\Delta_s^{\rm E}
\begin{bmatrix}
0\\
1
\end{bmatrix},
\\
\begin{bmatrix}
\hat V^{\rm I}[n+1]\\
\hat W^{\rm I}[n+1]
\end{bmatrix}
&\le
\mathcal M_k^{\rm I}
\begin{bmatrix}
\hat V^{\rm I}[n]\\
\hat W^{\rm I}[n]
\end{bmatrix}
+
\frac{\ell_2}{k\omega_k}
(\eta_k^{\rm I})^n\Delta_s^{\rm I}
\begin{bmatrix}
\ell_1\eta_k^{\rm I}/k\\
1
\end{bmatrix}.
\end{align}
\end{subequations}

Fix either update and omit its superscript. Let $\gamma_k:=(1+\rho(\mathcal M_k))/2$ and choose $\vartheta_k >0$ such that $[\vartheta_k ,1]\mathcal M_k \le\gamma_k[\vartheta_k ,1]$.
\footnote{Let $z_k^\top:=[1,1](\gamma_k I_2-\mathcal M_k)^{-1} =[z_{1,k},z_{2,k}]$. Since $\mathcal M_k\ge0$ and $\gamma_k>\rho(\mathcal M_k)$, the inverse is nonnegative with positive diagonal entries. Thus $z_{1,k},z_{2,k}>0$, and
$z_k^\top\mathcal M_k=\gamma_kz_k^\top-[1,1]$ gives the stated inequality with $\vartheta_k:=z_{1,k}/z_{2,k}$.}
With $U[n]:=\vartheta_k \hat V[n]+\hat W[n]$, \eqref{eq:app-transient-comparison} gives
\begin{equation}
U[n+1] \le \gamma_k U[n]+\upsilon_k\eta_k^n\Delta_s,
\label{eq:app-U-recursion}
\end{equation}
where $\upsilon_k^{\rm E}:=\ell_2/k$ and $\upsilon_k^{\rm I}:= \ell_2(k\omega_k)^{-1} (1+\vartheta_k \ell_1\eta_k^{\rm I}/k)$.
Iterating \eqref{eq:app-U-recursion} gives
\begin{equation}
U[n]\le\bar U[n]
:=\gamma_k^nU[0]
+\upsilon_k\Delta_sG_n(\gamma_k,\eta_k),
\label{eq:app-U-bound}
\end{equation}
where $U[0]=\vartheta_k(V[0]-V_{\infty,k})_+ +(W[0]-W_{\infty,k})_+$, $G_0(a,b):=0$, and $G_n(a,b):=\sum_{\ell=0}^{n-1}a^{n-1-\ell}b^\ell$ for $n\ge1$ and $a,b\in[0,1)$. For fixed $a,b\in[0,1)$ and $n\ge2$, $0\le G_n(a,b)\le n\max\{a,b\}^{n-1}\to0$.

Since $V[n]\le V_{\infty,k}+\hat V[n]\le V_{\infty,k}+\bar U[n]/\vartheta_k$ and $W[n]\le W_{\infty,k}+\hat W[n] \le W_{\infty,k}+\bar U[n]$, \eqref{eq:app-reconstruction} gives, for all $i\in\mathcal{N}$,
\begin{equation}
\begin{aligned}
\|x_i[n,k]-p_i s[n,k]\|
\le{} &
p_i\|H^{-1}\|V_{\infty,k}
+\|\mathsf T^{-1}\|W_{\infty,k}\\
&+
\left(
\frac{p_i\|H^{-1}\|}{\vartheta_k }
+\|\mathsf T^{-1}\|
\right)\bar U[n].
\end{aligned}
\label{eq:app-sync-final}
\end{equation}
For each fixed $k$ in the respective range, $\gamma_k,\eta_k\in[0,1)$. Since $n_{T,k}\to\infty$ as $T\to\infty$, \eqref{eq:app-U-bound} gives $\bar U[n_{T,k}]\to0$. Evaluating \eqref{eq:app-sync-final} at $n=n_{T,k}$ and using \eqref{eq:app-steady-bounds} bounds the synchronization term. Combining this bound with Lemma~\ref{lem_error} through the triangle inequality gives \eqref{eq:thm_1_E} and \eqref{eq:thm_1_I}. 

For later use, combining \eqref{eq:app-sync-final} with Lemma~\ref{lem_error2} gives the finite-iteration bound $\max_{i\in\mathcal N}\|x_i[n,k]-p_i\mathsf s(t_{n,k})\|\le e(n,k,x[0],s[0],\mathsf s(0))$ introduced in Section~\ref{subsection:trade-off}, where
\begin{equation}
\begin{aligned}
&e(n,k,x[0],s[0],\mathsf{s}(0))\\
&\quad:=
e_\infty(k)+C_k\bar U[n]\\
&\qquad+
p_{\max}\left(
\eta_k^n\|H^{-1}\|
\|H(s[0]-\mathsf{s}(0))\|
+\Phi_k[n]
\right),
\end{aligned}
\label{eq:app-error-bound}
\end{equation}
with $C_k:=p_{\max}\|H^{-1}\|/\vartheta_k +\|\mathsf T^{-1}\|$, $p_{\max}:=\max_{i\in\mathcal{N}}p_i$, $\bar U[n]$ given in \eqref{eq:app-U-bound}, $e_\infty(k)$ given in \eqref{eq:e-infty}, and $\Phi_k[n]$ denoting the second term on the right-hand side of the corresponding inequality in \eqref{eq:transient_discretization}.

Finally, we verify the uniform-in-time conclusion. Choose the auxiliary initial state $s[0]=q^\top_{\otimes m}x[0]$, so that $V[0]=0$. As $k\to\infty$, $\eta_k=1-\rho_{\mathsf{s}}/k+O(k^{-2})$ and $\gamma_k=1-\rho_{\mathsf{s}}/(2k)+O(k^{-2})$. The choice of $\vartheta_k$ in the above footnote satisfies $\vartheta_k=O(k)$ and $\vartheta_k^{-1}=O(k^{-1})$. Hence $U[0]$, $C_k$, $\Delta_s$, and $\bar\Delta_s$ are bounded uniformly for large $k$, while $\upsilon_k=O(k^{-1})$ and $e_\infty(k)=O(k^{-1})$. Using $G_n(a,b)\le n\max\{a,b\}^{n-1}$ for $n\ge1$ in \eqref{eq:app-U-bound} and \eqref{eq:app-error-bound} gives constants $C,c>0$, independent of $n$ and $k$, such that
\begin{equation*}
\begin{aligned}
&\max_{i\in\mathcal{N}}
\|x_i[n,k]-p_i\mathsf{s}(t_{n,k})\| \le
\frac{C}{k}+C(1+n/k)e^{-cn/k}
\end{aligned}
\end{equation*}
for all $n\ge0$ and all sufficiently large $k$. Since $n_{T,k}/k=T/\tau_{\mathsf{s}}+O(k^{-1})$ and \eqref{eq:vector-field-decay-full} gives $\|\mathsf{s}(t_{n_{T,k},k})-\mathsf{s}(T)\|=O(k^{-1})$ uniformly in $T$, for each $\epsilon>0$ there exist $k_0,T_0>0$ such that the approximation error is at most $\epsilon$ for all $k>k_0$ and $T\ge T_0$. The corresponding statement for $s[n_{T,k},k]$ follows from \eqref{eq:transient_discretization} and the same rounding estimate.

\subsection{Proof of Theorem~\ref{thm:iteration}}
\label{app:proof-theorem3}

Fix either update and omit its superscript. By \eqref{eq:eK} and \eqref{eq:app-error-bound}, we have $e_{\mathcal K}(n,k)\ge e_\infty(k)$. Conversely, for each fixed $k$ in the domain of $e_\infty$, choose $s[0]=\mathsf{s}(0)$ in the infimum in \eqref{eq:eK}. Then $U[0]$, $\Delta_s$, and $\bar\Delta_s$ are bounded uniformly over $\mathcal K_x$ and $\mathcal K_{\mathsf{s}}$ by compactness. Equations~\eqref{eq:app-U-bound} and \eqref{eq:transient_discretization} therefore give a constant $C>0$, independent of $n$, such that
\[
\begin{aligned}
0 & \le e_{\mathcal K}(n,k)-e_\infty(k)
\\ & \le C\left(\gamma_k^n+G_n(\gamma_k,\eta_k)+G_n(\eta_k,e^{-\rho_{\mathsf{s}}/k})\right).
\end{aligned}
\]
Since every term on the right tends to zero, $e_{\mathcal K}(n,k)\to e_\infty(k)$ as $n\to\infty$.
For $k>k_\epsilon$, we have $e_\infty(k)<\epsilon$, so $e_{\mathcal K}(n,k)\le\epsilon$ for all sufficiently large $n$ and $\mathbf N_\epsilon(k)<\infty$.

We now consider the two limiting cases.

\noindent\emph{1) $k\to\infty$.}

Let $\xi_1[0]=q^\top_{\otimes m}x[0]$ and $M_s:=\sup_{x[0]\in\mathcal K_x,\,\mathsf{s}(0)\in\mathcal K_{\mathsf{s}}}
\allowbreak\|H(\xi_1[0]-\mathsf{s}(0))\|$. By \eqref{eq:app-U-bound}, $\bar U[n]\ge\gamma_k^n\vartheta_k(V[0]-V_{\infty,k})_+$, where $V[0]=\|H(\xi_1[0]-s[0])\|$. Thus, the definition of $C_k$ gives
\[
\begin{aligned}
C_k\bar U[n]
&\ge p_{\max}\|H^{-1}\|\gamma_k^n \bigl(\|H(\xi_1[0]-s[0])\|-V_{\infty,k}\bigr)_+.
\end{aligned}
\]
Adding the term $p_{\max}\eta_k^n\|H^{-1}\|\allowbreak \|H(s[0]-\mathsf{s}(0))\|$ in \eqref{eq:app-error-bound}, and using $(a-v)_++b\ge(a+b-v)_+$ for $b\ge0$ together with the triangle inequality, gives a lower bound independent of $s[0]$. Hence, \eqref{eq:eK} yields
\begin{equation}
\begin{aligned}
e_{\mathcal K}(n,k)
&\ge e_\infty(k)+p_{\max}\|H^{-1}\|\\
&\quad{}\times\min\{\gamma_k^n,\eta_k^n\}
(M_s-V_{\infty,k})_+.
\end{aligned}
\label{eq:app-large-k-lower}
\end{equation}

Fix any $n\ge1$. As $k\to\infty$, $e_\infty(k)\to0$ and \eqref{eq:app-steady-bounds} gives $V_{\infty,k}\to0$. Also, $\mathcal M_k\to\operatorname{diag}(1,\mu_{\mathcal{W}})$, so $\gamma_k\to1$, while \eqref{eta_exp}--\eqref{eta_imp} give $\eta_k\to1$. Hence, the right-hand side of \eqref{eq:app-large-k-lower} converges to $p_{\max}\|H^{-1}\|M_s$. Under the first condition of Theorem~\ref{thm:iteration}, there exists $K_n>0$ such that $e_{\mathcal K}(n,k)>\epsilon$ for all $k>K_n$. Thus, $\mathbf N_\epsilon(k)>n$ for all $k>K_n$. Since $n$ is arbitrary, $\lim_{k\to\infty}\mathbf N_\epsilon(k)=\infty$.

\noindent\emph{2) $k\searrow k_\epsilon$.}

Let $\Delta_{\mathcal K_{\mathsf{s}}}:= \sup_{\mathsf{s}(0)\in\mathcal K_{\mathsf{s}}}\bar\Delta_s$. The second condition of Theorem~\ref{thm:iteration} gives $\Delta_{\mathcal K_{\mathsf{s}}}>0$. For each fixed $n\ge1$, write the transient term in \eqref{eq:transient_discretization} as $\Phi_k[n]=d_n(k)\bar\Delta_s$, where $d_n$ is continuous at $k_\epsilon$ and $d_n(k_\epsilon)>0$. This follows from $c_{\mathsf{s}}>0$, $G_n(a,b)\ge b^{n-1}>0$ for $a\ge0$, $b>0$, and the fact that $k_\epsilon$ lies inside the domain of $e_\infty$. Since $\Phi_k[n]$ is independent of $s[0]$, \eqref{eq:eK} and \eqref{eq:app-error-bound} give
\begin{equation}
e_{\mathcal K}(n,k)-e_\infty(k)
\ge
p_{\max}d_n(k)\Delta_{\mathcal K_{\mathsf{s}}}.
\label{eq:app-small-k-lower}
\end{equation}
Since $e_\infty(k)\to\epsilon$ as $k\searrow k_\epsilon$,
\eqref{eq:app-small-k-lower} gives
\[
\liminf_{k\searrow k_\epsilon}e_{\mathcal K}(n,k)
\ge
\epsilon+ p_{\max}d_n(k_\epsilon)\Delta_{\mathcal K_{\mathsf{s}}} >
\epsilon.
\]
Thus, $e_{\mathcal K}(n,k)>\epsilon$ and $\mathbf N_\epsilon(k)>n$ eventually as $k\searrow k_\epsilon$. Since $n$ is arbitrary, $\lim_{k\searrow k_\epsilon}\mathbf N_\epsilon(k)=\infty$, which completes the proof. 

\section{Proofs for the Applications}

\subsection{Proof of Proposition~\ref{prop:zeta_inverse}}
\label{prop:zeta_inverse_proof}

For each $\ell\in\mathcal{N}$, let $x_i^{(\ell)}[n]$ denote the $\ell$-th component of $x_i[n]$, and define $\chi^{(\ell)}[n]
:= \operatorname{col}\bigl(x_i^{(\ell)}[n]\bigr)_{i \in \mathcal{N}}$ and $\mathbf E_\ell := \mathbf e_\ell\mathbf e_\ell^\top$. 
Then, $\chi^{(\ell)}[n+1] = A_\ell\chi^{(\ell)}[n]+b_\ell$,
where $A_\ell:= \left(I_N+\frac{\Delta\mathbf E_\ell}{k}\right)^{-1}(I_N-\Delta\mathcal{L})$, $b_\ell:= \left(I_N+\frac{\Delta\mathbf E_\ell}{k}\right)^{-1}\frac{\Delta}{k}1_N$,
and $\Delta:=\operatorname{diag}(\delta_i)_{i\in\mathcal{N}}$. The matrix $A_\ell$ is nonnegative and irreducible. Its row sums are at most one, and its $\ell$-th row sum is strictly smaller than one. Hence $\rho(A_\ell)<1$ and $\chi^{(\ell)}[n]\to\chi^{(\ell)*}:=(\mathbf E_\ell+k\mathcal L)^{-1}1_N$. Since $\zeta^\top(\mathbf E_\ell+k\mathcal L) =\zeta_\ell\mathbf e_\ell^\top$ and $\zeta^\top1_N=1$, $\mathbf e_\ell^\top\chi^{(\ell)*}=1/\zeta_\ell$. Taking $\ell=i$ proves convergence; the nonnegative state matrix and positive input give positivity.

\subsection{Derivation for Example~\ref{ex:size_estimation_two_agent}}
\label{ex:size_est_two_proof}

For $N=2$ and $\delta_1=\delta_2=0.5$, write
$x[n+1]=A_kx[n]+b_k$, where $A_k=\operatorname{col}(\frac{k}{2k+1},\frac12)[1,1]$ and $b_k=\operatorname{col}((2k+1)^{-1},(2k)^{-1})$. The equilibrium is $x^\star=[2,\,2+1/k]^\top$, and $A_k$ has eigenvalues $0$ and $\lambda_k=(4k+1)/(4k+2)$. For $k>2$, the initial bounds in Example~\ref{ex:size_estimation_two_agent} give $|x_i[0]-x_i^\star|\le\bar N$ for $i=1,2$. Using $x[n]-x^\star=A_k^n(x[0]-x^\star)$ and $A_k^n=\lambda_k^{n-1}A_k$ for $n\ge1$ yields $|x_1[n]-2|\le2k\bar N\lambda_k^{n-1}/(2k+1)$ and $|x_2[n]-2|\le1/k+\bar N\lambda_k^{n-1}$. For $k>2$, the condition $\bar N\lambda_k^{n-1}<1/2-1/k$ is equivalent to $n>\Gamma_{0.5}(k)$ and makes both error bounds smaller than $0.5$. Thus nearest-integer rounding returns $2$ for $n\ge\mathbf N_{0.5}(k)$. The two limits follow from the expression for $\Gamma_{0.5}(k)$.

\section{Proofs of Lemmas~\ref{lem:discret_error} and~\ref{lem_error}}
\label{app:lemma-proofs}
\setcounter{lemma}{0}

\subsection{Proof of Lemma~\ref{lem:discret_error}}

The implicit update in \eqref{eq:emer_dyn} is uniquely solvable for $k>L_{\mathsf{s}}$, since $y\mapsto s[n]+f_{\mathsf{s}}(y)/k$ is a contraction. Fix $T>0$ and choose $M_T>0$ such that $\|f_{\mathsf{s}}(\mathsf{s}(t))\|\le M_T$ for $0\le t\le T+\tau_{\mathsf{s}}$. Such a bound exists because $f_{\mathsf{s}}$ is globally Lipschitz and $\mathsf{s}$ is continuous.

Let $e[n]:=s[n,k]-\mathsf{s}(t_{n,k})$ and define
\begin{align*}
r_n^{\rm E}
&:=\mathsf{s}(t_{n+1,k})-\mathsf{s}(t_{n,k})
-\frac1k f_{\mathsf{s}}(\mathsf{s}(t_{n,k})),\\
r_n^{\rm I}
&:=\mathsf{s}(t_{n+1,k})-\mathsf{s}(t_{n,k})
-\frac1k f_{\mathsf{s}}(\mathsf{s}(t_{n+1,k})).
\end{align*}
Using $\dot{\mathsf{s}}=f_{\mathsf{s}}(\mathsf{s})/\tau_{\mathsf{s}}$ and Lipschitz continuity gives
\[
\begin{aligned}
\|\mathsf H r_n^{\rm E}\|
\le{}&
\frac{L_{\mathsf{s}}\|\mathsf H\|}{\tau_{\mathsf{s}}^2}
\int_{t_{n,k}}^{t_{n+1,k}}
\int_{t_{n,k}}^t
\|f_{\mathsf{s}}(\mathsf{s}(u))\|\,du\,dt.
\end{aligned}
\]
For the implicit case, the inner integral is instead over $[t,t_{n+1,k}]$. Since $t_{n+1,k}-t_{n,k}=\tau_{\mathsf{s}}/k$,
both cases satisfy 
\begin{equation}\label{eq:app-shared-defect}
\|\mathsf H r_n^{\rm E}\|,\ \|\mathsf H r_n^{\rm I}\|
\le
\frac{L_{\mathsf{s}}\|\mathsf H\|}{2k^2}
\sup_{t_{n,k}\le t\le t_{n+1,k}}
\|f_{\mathsf{s}}(\mathsf{s}(t))\|.
\end{equation}

Taking $\mathsf H=I_m$ gives $\|r_n^{\rm E}\|,\ \|r_n^{\rm I}\| \le L_{\mathsf{s}}M_T/(2k^2)$ for $0\le n<n_{T,k}$ and $k\ge1$. Subtracting the identities defining $r_n^{\rm E}$ and $r_n^{\rm I}$ from \eqref{eq:emer_dyn} and using Lipschitz continuity together with $1+L_{\mathsf{s}}/k\le(1-L_{\mathsf{s}}/k)^{-1}$ gives, for either update and $k>L_{\mathsf{s}}$,
\[
\|e[n+1]\|
\le\frac{\|e[n]\|+L_{\mathsf{s}}M_T/(2k^2)}
{1-L_{\mathsf{s}}/k}.
\]
Since $e[0]=0$, iteration yields
\[
\|e[n]\|\le\frac{M_T}{2k}
\left((1-L_{\mathsf{s}}/k)^{-n}-1\right),
\qquad 0\le n\le n_{T,k}.
\]
This also holds when $L_{\mathsf{s}}=0$, since then $e[n]=0$. As $k\to\infty$, $(1-L_{\mathsf{s}}/k)^{-n_{T,k}}\to e^{L_{\mathsf{s}}T/\tau_{\mathsf{s}}}$, so $\|s[n_{T,k},k]-\mathsf{s}(t_{n_{T,k},k})\|\to0$. Since $t_{n_{T,k},k}\to T$, continuity of $\mathsf{s}$ proves the result.

\subsection{Proof of Lemma~\ref{lem_error}}
\label{app:lemma-contracting-proofs}

We first establish the following finite-iteration bound.

\begin{lemma}\label{lem_error2}
Under Assumption~\ref{assum:contractive}, let $s[n,k]$ and $\mathsf{s}(t)$ solve \eqref{eq:emer_dyn} and \eqref{eq:cont_emer} with arbitrary initial values $s[0],\mathsf{s}(0)\in\mathbb{R}^m$. For each $n\in\mathbb Z_{\ge0}$,
\begin{subequations}\label{eq:transient_discretization}
\begin{align}
& \|s[n,k]-\mathsf{s}(t_{n,k})\| \le 
 (\eta_k^{\rm E})^n\|H^{-1}\|\|H(s[0]-\mathsf{s}(0))\|  \notag \\
& \qquad + \frac{c_{\mathsf{s}}\bar\Delta_s}{k^2} G_n(\eta_k^{\rm E},e^{-\rho_{\mathsf{s}}/k})
+\frac{\beta_1^{\rm E}}{k-\bar k},
\label{eq:transient_discretization_E}\\
&\|s[n,k]-\mathsf{s}(t_{n,k})\| \le
(\eta_k^{\rm I})^n\|H^{-1}\|
\|H(s[0]-\mathsf{s}(0))\| \notag  \\
& \qquad + \frac{c_{\mathsf{s}}\eta_k^{\rm I}\bar\Delta_s}{k^2}G_n(\eta_k^{\rm I},e^{-\rho_{\mathsf{s}}/k}) +\frac{\beta_1^{\rm I}}{k},
\label{eq:transient_discretization_I}
\end{align}
\end{subequations}
with $k>\bar k$ in the explicit case and $k>0$ in the implicit case. Here, $\bar\Delta_s:=(\|H\mathsf{s}(0)\|-\|Hf_{\mathsf{s}}(0)\|/\rho_{\mathsf{s}})_+$, $c_{\mathsf{s}}:=L_{\mathsf{s}}^2\|H\|\|H^{-1}\|^2/2$, and $\eta_k^{\rm E},\eta_k^{\rm I}$ are given in \eqref{eta_exp}--\eqref{eta_imp}. The constants $\beta_1^{\rm E},\beta_1^{\rm I}>0$ are those in Theorem~\ref{thm:uniform}.
\end{lemma}

\begin{proof}
The implicit update is uniquely solvable for every $k>0$ by Lemma~\ref{lem:app-s-bound}. Let $y(t):=\|H\mathsf{s}(t)\|$. Assumption~\ref{assum:contractive} gives $D^+y(t)\le-\rho_{\mathsf{s}}y(t)/\tau_{\mathsf{s}} +\|Hf_{\mathsf{s}}(0)\|/\tau_{\mathsf{s}}$, where $D^+$ is the upper right Dini derivative. Scalar comparison and Lipschitz continuity give
\begin{align}
\|H\mathsf{s}(t)\|
&\le
\frac{\|Hf_{\mathsf{s}}(0)\|}{\rho_{\mathsf{s}}}
+e^{-\rho_{\mathsf{s}} t/\tau_{\mathsf{s}}}\bar\Delta_s,
\label{eq:continuous-absorbing-full}\\
\|f_{\mathsf{s}}(\mathsf{s}(t))\|
&\le
F_0+F_1e^{-\rho_{\mathsf{s}} t/\tau_{\mathsf{s}}}\bar\Delta_s,
\label{eq:vector-field-decay-full}
\end{align}
where $F_0:=\|f_{\mathsf{s}}(0)\| +L_{\mathsf{s}}\|H^{-1}\|\|Hf_{\mathsf{s}}(0)\|/\rho_{\mathsf{s}}$ and $F_1:=L_{\mathsf{s}}\|H^{-1}\|$.

For arbitrary $s[0]$ and $\mathsf{s}(0)$, use the error $e[n]$ and the terms $r_n^{\rm E},r_n^{\rm I}$ defined in the preceding subsection. Applying \eqref{eq:app-shared-defect} with $\mathsf H=H$ and \eqref{eq:vector-field-decay-full} gives
\begin{equation}\label{eq:local-defect-full}
\|Hr_n^{\rm E}\|,\ \|Hr_n^{\rm I}\|
\le
\frac{C_0}{k^2}
+\frac{C_1}{k^2}e^{-\rho_{\mathsf{s}}n/k}\bar\Delta_s,
\end{equation}
where $C_0:=L_{\mathsf{s}}\|H\|F_0/2$ and $C_1:=L_{\mathsf{s}}\|H\|F_1/2$.

Subtracting the identities defining $r_n^{\rm E}$ and
$r_n^{\rm I}$ from \eqref{eq:emer_dyn} and applying \eqref{eq:app-map-estimates} yields
\begin{align*}
\|He[n+1]\|
&\le
\eta_k^{\rm E}\|He[n]\|+\|Hr_n^{\rm E}\|,\\
\|He[n+1]\|
&\le
\eta_k^{\rm I}\|He[n]\|
+\eta_k^{\rm I}\|Hr_n^{\rm I}\|.
\end{align*}
Using \eqref{eq:local-defect-full} and iterating gives
\begin{subequations}
\begin{align}
\|He[n]\|
&\le
(\eta_k^{\rm E})^n\|He[0]\|
+\frac{C_0}{k^2}
\frac{1-(\eta_k^{\rm E})^n}{1-\eta_k^{\rm E}}
\notag\\
&\quad+
\frac{C_1\bar\Delta_s}{k^2}
G_n(\eta_k^{\rm E},e^{-\rho_{\mathsf{s}}/k}),
\label{eq:sharp-E-H-full}\\
\|He[n]\|
&\le
(\eta_k^{\rm I})^n\|He[0]\|
+\frac{C_0\eta_k^{\rm I}}{k^2}
\frac{1-(\eta_k^{\rm I})^n}{1-\eta_k^{\rm I}}
\notag\\
&\quad+
\frac{C_1\eta_k^{\rm I}\bar\Delta_s}{k^2}
G_n(\eta_k^{\rm I},e^{-\rho_{\mathsf{s}}/k}).
\label{eq:sharp-I-H-full}
\end{align}
\end{subequations}

By \eqref{eq:one-minus-eta-E-full} and $\eta_k^{\rm I}/(1-\eta_k^{\rm I})=k/\rho_{\mathsf{s}}$, the terms containing $C_0$ are bounded by $\beta_0/(k-\bar k)$ and $\beta_0/k$, respectively, after multiplication by $\|H^{-1}\|$, where $\beta_0:=C_0\|H^{-1}\|/\rho_{\mathsf{s}}$. Moreover, $c_{\mathsf{s}}=C_1\|H^{-1}\| =L_{\mathsf{s}}^2\|H\|\|H^{-1}\|^2/2$. Choose positive $\beta_1^{\rm E},\beta_1^{\rm I}$ with $\beta_1^{\rm E},\beta_1^{\rm I}\ge\beta_0+F_0$. Since $\|e[n]\|\le\|H^{-1}\|\|He[n]\|$, these choices prove Lemma~\ref{lem_error2}.
\end{proof}

For Lemma~\ref{lem_error}, fix $k$ in the stated range and set $n=n_{T,k}$ in \eqref{eq:sharp-E-H-full}--\eqref{eq:sharp-I-H-full}. The initial-condition terms and the $G_n$ terms vanish as $T\to\infty$. Since $|t_{n_{T,k},k}-T|\le\tau_{\mathsf{s}}/k$, \eqref{eq:vector-field-decay-full} gives
\[
\begin{aligned}
&\|\mathsf{s}(t_{n_{T,k},k})-\mathsf{s}(T)\|
\le\frac{F_0}{k} +\frac{F_1\bar\Delta_s}{k}
e^{-\rho_{\mathsf{s}}\min\{T,t_{n_{T,k},k}\}/\tau_{\mathsf{s}}}.
\end{aligned}
\]
The triangle inequality therefore gives, respectively,
\begin{align*}
&\limsup_{T\to\infty}
\|s[n_{T,k},k]-\mathsf{s}(T)\|\le
\frac{\beta_0}{k-\bar k}+\frac{F_0}{k}
\le\frac{\beta_1^{\rm E}}{k-\bar k},\\
&\limsup_{T\to\infty}
\|s[n_{T,k},k]-\mathsf{s}(T)\|\le
\frac{\beta_0+F_0}{k}
\le\frac{\beta_1^{\rm I}}{k},
\end{align*}
which proves Lemma~\ref{lem_error}.

\section{The Cases $\rho(\mathcal{W})\ne1$}
\label{app:spectral-cases}

We give the details for the two other spectral regimes discussed
in Section~\ref{section:2}.

\begin{proposition}\label{prop:spectral-cases}
Let Assumptions~\ref{assum:graph} and~\ref{assum:b,c} hold, and suppose that each $f_i$ is globally Lipschitz. Set $\lambda:=\rho(\mathcal{W})$.
\begin{enumerate}
\item[(i)] If $\lambda<1$, then, for all sufficiently large $k$, both updates converge globally exponentially to the same unique equilibrium $x_k^*$, and $\|x_k^*\|\to0$ as $k\to\infty$.

\item[(ii)] If $\lambda>1$, then, for every sufficiently large $k$, each update has an unbounded trajectory. When $F(0)=0$, such trajectories can start arbitrarily near the origin.
\end{enumerate}
\end{proposition}

\begin{proof}
\noindent\emph{(i) $\lambda<1$.}
Let $p>0$ satisfy $\mathcal{W}p=\lambda p$ and define $\|x\|_p:=\max_{i\in\mathcal{N}}\|x_i\|/p_i$. Since $\mathcal{W}$ is nonnegative and $A$ is diagonal, $\|\mathcal{W}_{\otimes m}\|_p=\lambda$ and $A_{\otimes m}F$ is $\ell$-Lipschitz in this norm, where $\ell:=L\|A\|$. The explicit transition map therefore has Lipschitz constant at most $\lambda+\ell/k$. For $k>\ell$, the implicit next-state equation is uniquely solvable, and comparison of two solutions gives the bound
$\lambda/(1-\ell/k)$. Thus, for $k>\ell/(1-\lambda)$, both maps are contractions and have unique globally exponentially attracting fixed points. These coincide because both solve $x_k^*=\mathcal{W}_{\otimes m}x_k^*+A_{\otimes m}F(x_k^*)/k$. With $c:=\|A_{\otimes m}F(0)\|_p$, this equation gives 
\begin{equation*}
\|x_k^*\|
\le\|p\|\|x_k^*\|_p
\le\frac{\|p\|c}{k(1-\lambda)-\ell}
\longrightarrow0
\end{equation*}
as $k\to\infty$.

\noindent\emph{(ii) $\lambda>1$.}
Primitivity gives positive $p,q$ with $q^\top p=1$ and a decomposition
\begin{equation*}
\mathcal{W}
=[p\ R]
\begin{bmatrix}\lambda&0\\0&\Lambda\end{bmatrix}
\begin{bmatrix}q^\top\\Z^\top\end{bmatrix},
\qquad \rho(\Lambda)<\lambda.
\end{equation*}
Choose a nonsingular $\mathsf S$ with $\mu:=\|\mathsf S\Lambda\mathsf S^{-1}\|<\lambda$, and set $u=q^\top_{\otimes m}x$ and $v=(\mathsf S Z^\top)_{\otimes m}x$. Writing $u^+,v^+$ for the next-iteration values, the explicit update becomes
\begin{align*}
u^+&=\lambda u+\frac{\phi(u,v)}{k},\\
v^+&=(\mathsf S\Lambda\mathsf S^{-1})_{\otimes m}v
+\frac{\psi(u,v)}{k},
\end{align*}
where
\begin{equation*}
\begin{aligned}
\begin{bmatrix}\phi(u,v)\\\psi(u,v)\end{bmatrix}
&=
\begin{bmatrix}q^\top A\\\mathsf S Z^\top A\end{bmatrix}_{\otimes m}
\times
F\bigl(p_{\otimes m}u+(R\mathsf S^{-1})_{\otimes m}v\bigr).
\end{aligned}
\end{equation*}
The implicit update evaluates $\phi,\psi$ at $(u^+,v^+)$ instead. Global Lipschitz continuity gives $K>0$ and $C\ge0$ such that
\begin{equation}\label{eq:spectral-large-growth}
\|\phi(u,v)\|,\ \|\psi(u,v)\|
\le K(\|u\|+\|v\|)+C.
\end{equation}
When $F(0)=0$, take $C=0$.

We seek initial states for which $\|u\|$ grows geometrically while $\|v\|\le\|u\|$ is preserved. Consider $\mathcal C:=\{(u,v):\|u\|\ge C/K,\ \|v\|\le\|u\|\}$. On $\mathcal C$, write $r=\|u\|$. For the explicit update, \eqref{eq:spectral-large-growth} bounds each nonlinear contribution by $3Kr/k$. For the implicit update, require $k>L\|A\|$ for well-posedness. With $z^+=\|u^+\|+\|v^+\|$, \eqref{eq:spectral-large-growth} gives
\begin{equation*}
z^+\le(\lambda+\mu)r+\frac{2K}{k}z^++\frac{2C}{k}.
\end{equation*}
Since $C\le Kr$, for $k\ge4K$ we obtain
$z^+\le2(\lambda+\mu+1)r$ and hence $Kz^++C\le Dr$,
where $D:=K(2\lambda+2\mu+3)>3K$.
Thus \eqref{eq:spectral-large-growth} gives, for both updates,
\begin{equation*}
\|u^+\|\ge(\lambda-D/k)r,\qquad
\|v^+\|\le(\mu+D/k)r.
\end{equation*} 
For sufficiently large $k$, $\lambda-D/k>\max\{1,\mu+D/k\}$. Hence $\mathcal C$ is forward invariant, and every trajectory starting there with $u[0]\ne0$ has $\|u[n]\|$ growing geometrically. Since $u=q^\top_{\otimes m}x$, each such trajectory is unbounded. If $F(0)=0$, then $C=0$, so these initial states can be chosen arbitrarily close to the origin.
\end{proof}

\end{document}